\documentclass[a4paper,fleqn,usenatbib]{mnras}
\usepackage{threeparttable}
\usepackage{txfonts}

\usepackage[T1]{fontenc}
\usepackage{ae,aecompl}
\usepackage{rotating}
\usepackage{tablefootnote}
\usepackage{xcolor}
\usepackage{graphicx}	
\usepackage{amssymb}	

\title[Dwarf kinematics in a group accreted to Virgo]{On the accretion of a new group of galaxies onto Virgo: I. Internal kinematics of nine in-falling dEs}

\author[B. Bidaran et al.]{
Bahar Bidaran,$^{1}$\thanks{E-mail: bidaran@uni-heidelberg.de (BB)}
Anna Pasquali,$^{1}$
Thorsten Lisker,$^{1}$
Lodovico Coccato,$^{2}$
\newauthor
Jesus Falc\'on-Barroso,$^{3,4}$
Glenn van de Ven,$^{5}$
Reynier Peletier,$^{6}$
Eric Emsellem,$^{2}$
\newauthor
Eva K. Grebel,$^{1}$
Francesco La Barbera,$^{7}$
Joachim Janz,$^{8,9}$
Agnieszka Sybilska,$^{10}$
\newauthor
Rukmani Vijayaraghavan,$^{11}$
John Gallagher III,$^{12}$
Dimitri A. Gadotti$^{2}$
\\
\\
$^{1}$Astronomisches Rechen-Institut, Zentrum f\"ur Astronomie der Universit\"at Heidelberg, M\"onchhofstra\ss e 12-14, 69120 Heidelberg, Germany\\
$^{2}$European Southern Observatory, Karl-Schwarzschild-Str. 2, D-85748 Garching, Germany\\
$^{3}$Instituto de Astrof\'isica de Canarias, Calle V\'ia L\'actea s/n, E-38205 La Laguna, Tenerife, Spain\\
$^{4}$Departamento de Astrof\'isica, Universidad de La Laguna, E-38200 La Laguna, Tenerife, Spain \\
$^{5}$Department of Astrophysics, University of Vienna, T\"urkenschanzstrasse 17, 1180 Vienna, Austria\\
$^{6}$Kapteyn Astronomical Institute, University of Groningen, Postbus 800, 9700 AV Groningen, the Netherlands\\
$^{7}$INAF-Osservatorio Astronomico di Capodimonte, sal. Moiariello 16, Napoli, 80131, Italy\\
$^{8}$Space Science and Astronomy, P.O. Box 3000, FI-90014 University of Oulu, Finland\\
$^{9}$Finnish Centre of Astronomy with ESO (FINCA), Vesilinnantie 5, FI-20014 University of Turku, Finland\\
, University of Turku, Väisäläntie 20, FI-21500 Piikkiö, Finland\\
$^{10}$Sybilla Technologies, Torunska 59, 85-023 Bydgoszcz, Poland\\
$^{11}$Department of Astronomy, University of Virginia, 530 McCormick Road, Charlottesville, VA 22904, USA\\
$^{12}$Department of Astronomy, University of Wisconsin- Madison, 475 N. Charter Street, Madison, WI 53076-1582, USA\\
}

\begin{document}
\pagerange{\pageref{firstpage}--\pageref{lastpage}}
\volume{---}
\pubyear{---}

\label{firstpage}

\maketitle
\begin{abstract}
Galaxy environment has been shown to play an important role in transforming late-type, star-forming galaxies to quiescent spheroids. This transformation is expected to be more severe for low-mass galaxies (M < $10^{10}$ \(M_\odot\)) in dense galaxy groups and clusters, mostly due to the influence of their past host halos (also known as pre-processing) and their present-day environments. For the first time, in this study, we investigate a sample of nine early-type dwarf galaxies (dEs) that were accreted as a likely bound group onto the Virgo galaxy cluster about 2-3 Gyr ago. Considering this special condition, these nine dEs may provide a test bed for distinguishing between the influence of the Virgo galaxy cluster and the effects of the previous host halo on their current properties. Specifically, we use VLT/MUSE integral-field unit spectra to derive their kinematics and specific angular momentum ($\lambda_{R}$) profiles. We observe a spread in the $\lambda_{R}$ profiles of our sample dEs, finding that the $\lambda_{R}$ profiles of half of them are as high as those of low-mass field galaxies. The remaining dEs exhibit $\lambda_{R}$ profiles as low as those of Virgo dEs that were likely accreted longer ago. Moreover, we detect nebular emission in one dE with a gas velocity offset suggesting ongoing gas stripping in Virgo. We suggest that the low-$\lambda_{R}$ dEs in our sample were processed by their previous host halo, prior to their infall to Virgo, and that the high-$\lambda_{R}$ dEs may be experiencing ram pressure stripping in Virgo.
\end{abstract}


\section{Introduction}\label{introduction}

Of the many different types of galaxies in the Universe, dwarf elliptical galaxies (dEs) are among the most numerous ones by number \citep{1985Sandage, 1988Binggeli, 2002Trentham}. The probability of observing dEs in high-density environments, such as groups and clusters, is higher than in the field \citep{1988Binggeli,2006Boselli, 2012Geha}. Studies of the morphology within one effective radius of these galaxies have shown that their structure is not as uniform as previously thought, since some show different kinds of substructures, for instance nuclei, spiral arms, and disks \cite[e.g.,][]{2006Lisker}. The observed sub-structures such as disk components in bright dEs \citep{1991Binggeli,2000Jerjen,2002Barazza,2003Graham,2003DeRijcke,2006Lisker, 2007Lisker}, the projected axial ratio distribution of cluster dEs \citep{2007Lisker}, as well as their prolonged star formation history \citep{2008Michielsen,  2010Paudel, 2011Koleva} and complex internal dynamics \citep{2003Geha}, favor two possible formation channels: through merging \citep{2005deRijcke} or through transformation of star-forming progenitors into quiescent dwarf galaxies due to environmental effects \cite[see e.g.,][]{1985Kormendy, 1988Binggeli, 1996Moore, 2010Geha, 2011Toloba, 2012Janz, 2012Kormendy, 2013Rys, 2013Lisker, 2014Boselli, 2014Penny, 2015Bialas, 2016Aguerri, 2018Hwang}. 

In fact, upon accretion onto a cluster or a group \citep{Binggeli1987, conselice2001}, galaxies undergo transformations as the result of their interactions with the higher-density environment and members of the host halo. At intermediate cluster-centric distances, ram pressure stripping can efficiently remove gas and suppress star formation relatively fast in less massive dwarf progenitors \citep{1972Gunn,1983Lin, 1985Kormendy, 2006Boselli, 2007Chung, 2014Bosellib}.
The gravitational interactions between the infalling galaxy and the cluster potential well during their orbits \citep{2015Smith} as well as close encounters with other cluster members
can trigger morphological and kinematic disturbances (e.g., destroying the disk substructure, reducing the rotation of the stellar component, and removing dark matter and stellar mass) within several orbital periods \citep{2006Boselli, 2006Lisker, 2005Mastropietro, 2013Lisker, 2015Bialas, 2015FalconBarroso}. Such environmental mechanisms are more severe for low-mass galaxies, mainly due to their shallow potential wells. For instance, as shown in the kinematic studies by \cite{2011Toloba} and \citet{2015Toloba}, the dense core of the Virgo cluster is mostly populated by dEs that are preferentially pressure-supported, while most dEs in the cluster outskirts show a higher degrees of rotation \citep[see also][]{2014Bosellib}.

\begin{figure}
\includegraphics[scale=0.55]{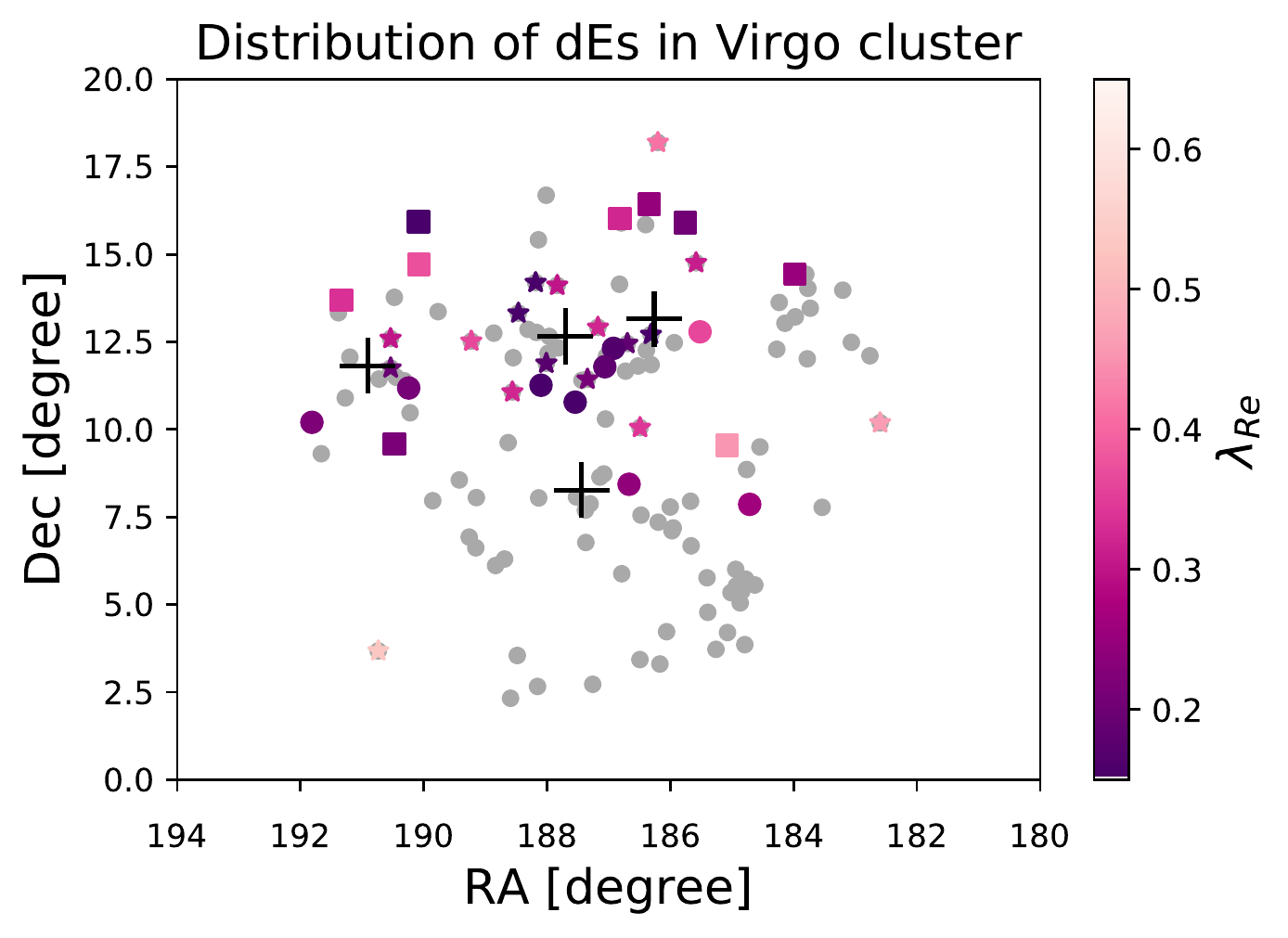}
\includegraphics[scale=0.55]{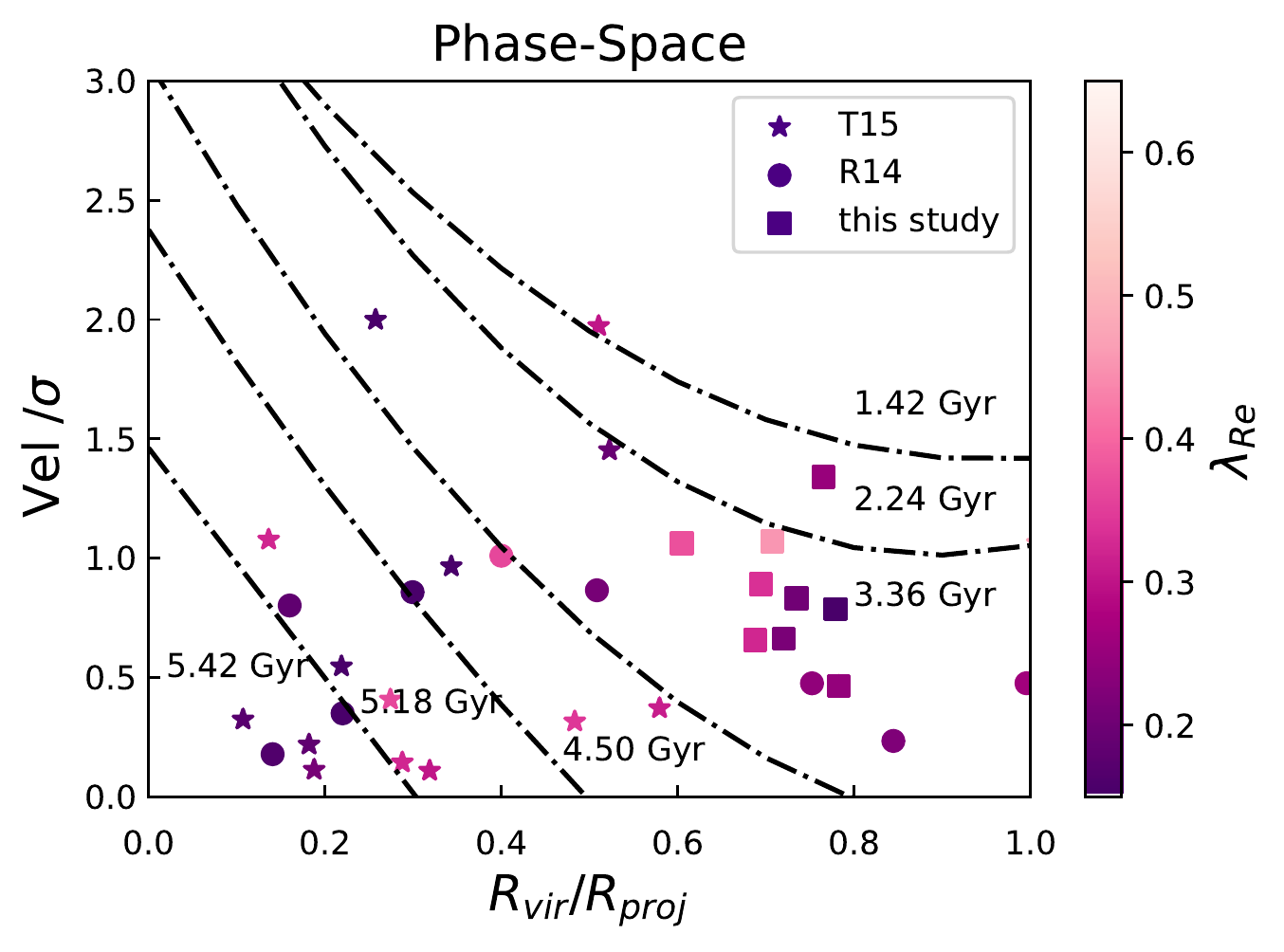}
\caption{\textit{Top panel}: Spatial distribution of dEs in the Virgo cluster. In both plots, squares represent the distribution of our sample of dEs colour-coded based on their $\lambda_{Re}$ (Section \ref{lambda_R}). The dEs in the Rys et al. (2014) sample (R14) are denoted with circles and colours that are scaled accordingly. The same applies to the dEs in the Toloba et al. (2015) sample (T15) that are denoted with star symbols. Gray circles represent the distribution of Virgo dEs with $-17\geq M_{r}> -18 $. Positions of Virgo giant early-type galaxies (from left to right: M60, M87, M49, M86) are marked with black crosses. \textit{Bottom panel}: projected phase-space distribution of our sample of dEs and those of R14. The different labelled zones of the diagram, taken from Pasquali et al. (2019) and Smith et al. (2019), represent different average infall times to the host halo (shown here for the Virgo cluster).}
\label{Virgo_distribution}
\end{figure}

\setlength{\tabcolsep}{3.pt}
\begin{table*}
\caption{\label{Observe} List of targets}
\centering
\begin{tabular}{c c l l c c c c c c c}
\hline
Object      & type    & $\alpha$ (J2000)    &   $\delta$ (J2000)  &  V$_{\rm recession}^{a}$ [km\,s$^{-1}$]  & \textit{R}$_{\rm e}^{b}$ \rm [arcsec] & \textit{M}$_{\rm r}^{b}$ \rm [mag] & \textit{g-r}$^{a}$ \rm [mag] &  $\epsilon ^{b}$  & log (\textit{M}$_{\star}$ [M$_\odot$]) & TET [hour]\\
\hline
\hline
VCC 0170 & dE(bc) & 12 15 56.30 & +14\ 25\ 59.2 & 1415.0 & $31\farcs57$ & -17.62 & 0.59  & 0.34& 9.14 & 4\\
VCC 0407 & dE(di) & 12 20 18.80 & +09\ 32\ 43.1 & 1876.7 & $18\farcs38$ & -17.37 & 0.61  & 0.43 & 9.06 &2\\
VCC 0608 & dE(di) & 12 23 01.70 & +15\ 54\ 20.2 & 1819.7 & $25\farcs77$ & -17.58 & 0.60  & 0.35 & 9.14 &5\\
VCC 0794 & dE(nN) & 12 25 21.61 & +16\ 25\ 46.9 & 1672.8 & $37\farcs33$ & -17.29 & 0.61  & 0.65 & 9.02 &3\\
VCC 0990 & dE(di) & 12 27 16.93 & +16\ 01\ 28.1 & 1717.8 & $10\farcs31$ & -17.43 & 0.62  & 0.30 & 9.10 &3\\
VCC 1833 & ---    & 12 40 19.70 & +15\ 56\ 07.1 & 1705.8 & $8\farcs52$  & -17.44 & 0.61  & 0.19 & 9.09 &1.5\\
VCC 1836 & dE(di) & 12 40 19.50 & +14\ 42\ 54.0 & 2002.6 & $42\farcs27$ & -17.45 & 0.58  & 0.66 & 9.06& 4\\ 
VCC 1896 & dE(di) & 12 41 54.60 & +09\ 35\ 04.9 & 1885.7 & $14\farcs98$ & -17.04 & 0.62  & 0.05 & 8.94& 3\\
VCC 2019 & dE(di) & 12 45 20.40 & +13\ 41\ 34.1 & 1819.7 & $18\farcs60$ & -17.65 & 0.63  & 0.22 & 9.16& 2\\
\hline
\end{tabular}\\
Columns are: Name of target, morphological type, RA and DEC, recession velocity (V$_{\rm recession}$), effective radius ($R_e$) (half-light major axis), r-band absolute magnitude, \textit{g-r} colour measured at 1$R_{\rm e}$, ellipticity at 1$R_{\rm e}$, stellar mass ($M_\star$), total exposure time (TET).\\
a:\,SDSS DR15 \citep{2019Aguado},
b:\,\cite{2006Lisker, 2007Lisker}
\\

\end{table*}

Galaxy rotation can be quantified in terms of specific angular momentum ($\lambda_{R}$). $\lambda_{R}$ is a proxy for the projected angular momentum of a galaxy, which is often described as an indispensable metric in studies of galaxy evolution \citep{2007Emsellem, 2009Jesseit, 2011Emsellem}. Through the evolution of a given galaxy, this parameter is affected by gas accretion and star formation activity as well as by the interaction of the galaxy with its surrounding environment  \citep[see][]{ 2009Jesseit, 2014Naab, 2016Yozin, 2017Penoyre, 2020arXiv200408598W}. This makes $\lambda_{R}$ a valuable, but degenerate, metric for a better understanding of galaxy formation and evolution. Investigations of \citet[hereafter R14]{2014Rys} show that dEs in Virgo (as well as two dEs in the field) tend to show flat $\lambda_{R}$ profiles up to one effective radius. A similar rather flat profile is also observed for dwarf galaxies in Fornax (see Scott et al. in preparation). As suggested by R14, if host halos, such as clusters, are in charge of any transformation in the kinematics of dE progenitors, then at fixed stellar mass, cluster dEs with different infall time are expected to exhibit different internal kinematics. In this picture, dEs with a more recent infall time are expected to show internal kinematics intermediate between low-mass field galaxies and cluster dEs accreted at earlier times. In fact, studies of \cite{2011Toloba} show that in the outer parts of the Virgo cluster and at a fixed range of luminosity, dEs tend to show similar rotation curves to late-type galaxies. This is particularly expected in a dynamically young galaxy cluster with ongoing processes of assembly and a diverse, yet complete population of galaxies, such as Virgo \citep{2006Boselli, 2018Boselli}. In contrast, if dEs are already systems that formed with low and flat $\lambda_{R}$ profiles \citep{2017Wheeler}, their environment or infall time will not significantly modify their specific angular momentum. Observations of dEs with low degrees of rotation in the field by \cite{2017Janz} also challenge the idea that only tidal interactions in clusters can decrease $\lambda_{R}$.

A recently accreted group of galaxies in the Virgo cluster was discovered by \cite{lisker2018}, based on the clustering of nine dEs with $-17\geq M_{r}> -18 $ in a particular region of the observer's phase-space plane. These relatively young dEs (based on their NUV-r colour) are found at a projected cluster-centric distance of $\approx$ 1.5 Mpc with respect to M87, and move with line-of-sight velocities of $\approx$ 700 km\,s$^{-1}$ relative to Virgo. Most of them feature substructures, such as disks and spiral arms \citep{2006Lisker,2006LiskerII, lisker2018}. They show a sparsely-defined ellipsoidal distribution azimuthally around M87 in Virgo, confined to the northern part of the cluster (see top panel of Fig. \ref{Virgo_distribution}). N-body simulations of infalling galaxy groups support their particular phase-space distribution as the result of a galaxy group accretion onto Virgo. According to simulations, this accretion event has taken place 2 to 3 Gyr ago, along the observer's line-of-sight \citep{2015Vijayaraghavan}. Such a picture receives further support by studies of \cite{2019Pasquali} and \cite{2019Smith}, who showed that cluster galaxies occupy different zones in the phase-space diagram with respect to their average infall time to cluster. Based on the phase-space locus of these nine dEs (colored squares in the lower panel of Fig. \ref{Virgo_distribution}), it can be clearly seen that they all share similar, recent infall times to Virgo. Since members of this in-falling group are in the initial phase of accretion, the assembly history and dynamical characteristics of their previous environment (i.e., the parent group) should still be preserved in their kinematics and stellar populations. This newly accreted group of dEs to Virgo provides a unique opportunity for investigating the role of a cluster environment in the internal dynamical evolution of dE progenitors, particularly during the early stages of accretion. In addition to the Virgo cluster, environmental mechanisms in their previous host halo (such as ram pressure stripping, tidal interactions and starvation) have possibly affected the evolution of these nine dEs before their accretion onto Virgo. This represents the condition that is commonly known as pre-processing \citep[e.g.,][]{1989Gallagher, 2004Fujita, 2004Mihos, 2017Joshi, 2018Han, 2019Joshi}. To this end, studying this group of dEs can, also address the role of pre-processing in shaping present-day properties of the "typical" dE population in clusters, as suggested by \cite{2014Toloba} and \cite{2017Sybilska}. 

Motivated by this, we obtained IFU (Integral Field Units) data for the nine dEs in this group using the MUSE (Multi-Unit Spectroscopic Explorer) instrument at the Very Large Telescope (VLT). We have performed an extensive analysis of the kinematics, dynamics, and stellar populations of this sample. The results will be presented in a series of papers. In this first paper, we present our kinematic analysis of this sample of nine dEs.

This paper is organized as follows: In Section \ref{observation}, we describe our data set. In Section \ref{methods} the methods and approaches that are used in this investigation are discussed in detail. Section \ref{Results} shows the kinematic maps and the specific angular momentum ($\lambda_{R}$) profile of each dE in this sample. In this section we also compare the kinematic properties of our sample with other dEs in Virgo and low-mass galaxies in the field. We discuss our results in Section \ref{discussion}. Our conclusions follow in Section \ref{conclusion}.

\section{Observations and data reduction}\label{observation}

Our sample consists of nine dEs of a galaxy group that was accreted recently to the Virgo cluster, discovered by \cite{lisker2018}. The name of each target, its type, coordinates, recession velocity, effective radii from r-band photometry images, r-band absolute magnitude, colour, ellipticity, and total exposure time (TET), are summarized in Table \ref{Observe}. All the values are taken from \cite{2006Lisker}, except for the recession velocity which are taken from SDSS DR15 \citet{2019Aguado}. The reported values of $m_{r}$ (r-band apparent magnitude) and colour were corrected for Galactic extinction by \cite{2006Lisker} and \cite{2008Janz, 2009Janz}. We estimate the stellar mass of our sample of dEs to be 8.9 $\leq$ log(\textit{M}$_{\star}$ [M$_\odot$]) $\leq$ 9.2\, (also reported in Table \ref{Observe}), based on the colour-to-(M/L) conversion (M and L denote mass and light of system, respectively) introduced by \cite{2003bell} and (g-r) colours (reported in Table \ref{Observe}) using a distance modulus of m-M = 31.09 mag, where m denotes the apparent and M the absolute magnitude.

In Fig. \ref{Virgo_distribution} we show the spatial distribution of these nine dEs (top panel) as well as their location in the projected phase-space diagram (lower panel).
In the top panel, other Virgo dEs with similar r-band absolute magnitude are shown with gray circles \citep{2006Lisker, 2008Janz, 2009Janz}. The positions of four giant elliptical galaxies in Virgo are marked with black crosses. We also over-plotted the distribution of our comparison sample from R14 and T15 in both panels, marked with colour-coded circles and stars, respectively. The colour bars show the values of the specific angular momentum within one effective radius (see Section \ref{lambda_R}). In the lower panel, dash-dotted lines distinguish different zones of different average infall times in the observed phase-space diagram, which were adapted from \cite{2019Pasquali} and \cite{2019Smith}. The position of all dEs, their projected distances, and their line of sight velocities are from \cite{2006Lisker}. The virial radius of Virgo adopted here is 1.55 Mpc \citep{1999McLaughlin}.

The targets were observed with the MUSE following a science verification proposal in the period of December 2016 to February 2017 and February 2018 to July 2018 (P98, ESO programmes 098.B-0619 and 0100.B-0573; PI: Lisker). MUSE is an integral-field spectrograph mounted on VLT of the European Southern Observatory, located on Paranal, Chile. MUSE consists of 24 spectrographs (IFUs). With a pixel scale of 0.2 arcsec/pixel, this instrument provides a field of view (FOV) of ${1}' \times {1}'$ in wide-field mode (WFM). MUSE delivers 90000 sampled spectra with a resolving power of $R=$3000 and a final sampling of 1.25 \AA/pix over the range of 4500-9300 \AA. MUSE's average instrumental resolution has a full width at half maximum (FWHM) = $2.51$ \AA\, \citep{2010Bacon}. 
 
During the observing runs the average seeing was nearly constant with a FWHM of about 1.6$\arcsec$. The reduction of the data and the construction of the final calibrated data cubes were carried out within the ESO REFLEX environment \citep{2013Freudling} using the standard MUSE Data Reduction Software (version 2.4.2) \citep{2012Weilbacher, 2016Weilbacher}. Sky residuals were removed from the final data cube by using MUSE-ZAP (v.2.2) \citep{2016Soto}.

In this paper, we will compare the kinematic properties of our sample of dEs with two available samples in the literature. Our Virgo control sample consists of nine dEs investigated by \citet[hereafter R14 sample]{2014Rys} and 21 dEs investigated by \citet[hereafter T15 sample]{2015Toloba}. While the original sample of T15 consists of 39 dEs, we only selected those that are consistent with the mass range of our sample. The galaxies of our field control sample (hereafter CALIFA field galaxies) are part of the CALIFA sample \citep{2012A&A...538A...8S, 2014Walcher, 2017FalconBarroso} with $M_{\rm \star} < 5 \times 10^{9}$ M$_{\odot }$, consistent with the stellar mass range of the dEs in our sample and of the R14 and T15 sample (for more information see Appendix \ref{Califa_galaxies_appendix}). Our field control sample consists of late-type spiral and irregular galaxies. 


\section{Methods}\label{methods}
\subsection{Measurements of stellar kinematics}

\begin{figure*}
\centering
\includegraphics[scale= 0.6]{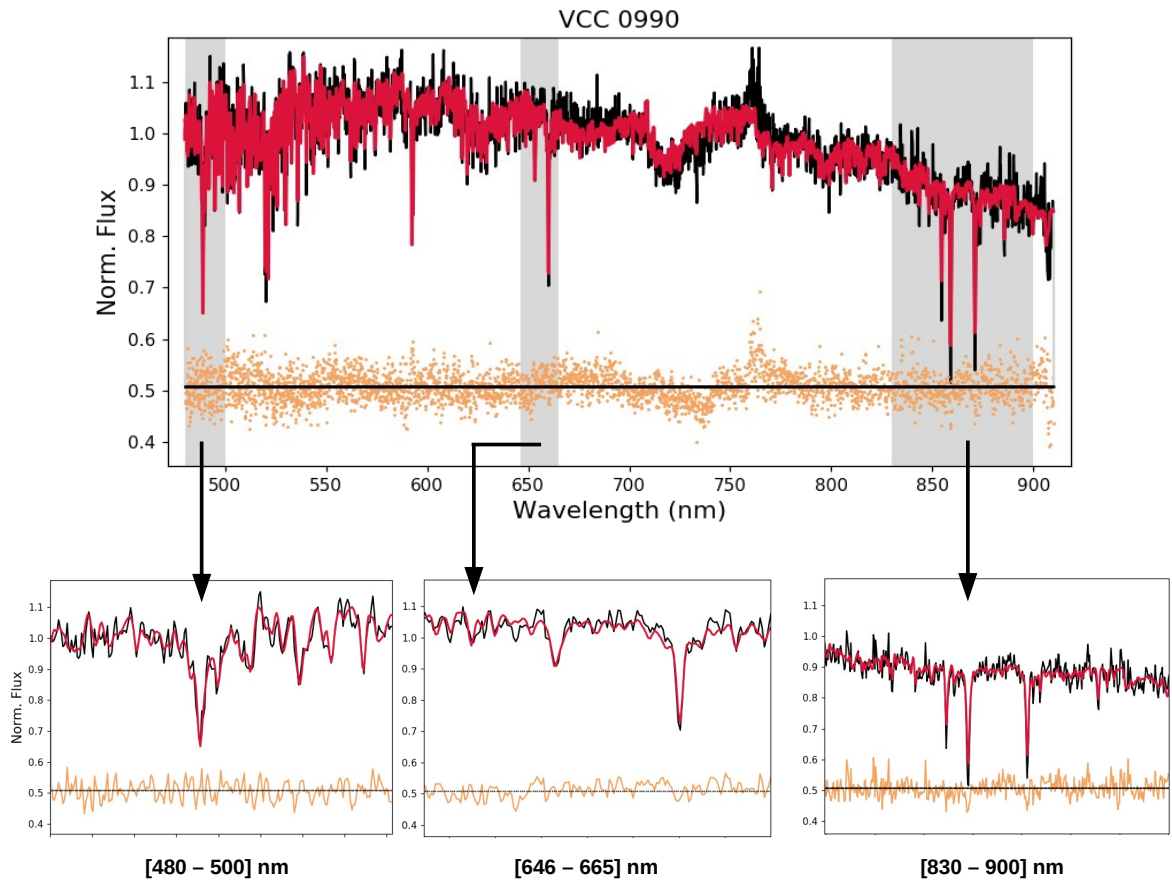}
\caption{Example of a pPXF fit of a central bin of VCC0990. The top panel represents the full range of the observed spectrum, while in the lower panels, a zoomed-in view is provided for a better illustration of the fitting accuracy and high quality of the data. In all the panels, the observed spectrum is shown in black, while the best pPXF fit is plotted in red. The residuals between the observed spectrum and its best fit are plotted in orange shifted up by 0.5 for legibility.}
\label{fitting_quality}
\end{figure*}

Our sample consists of low-surface-brightness objects. To conduct accurate measurements of their kinematic properties, the signal-to-noise ratio (SNR) in each data cube needed to be increased to a minimum threshold. While considering only spaxels with SNR $>3$, we increased the final SNR for each galaxy by binning spaxels through the Voronoi binning method developed by \cite{cappellari2003}. This routine is an optimized fitting algorithm that spatially bins data in a way such that each final bin achieves the required SNR. While setting low values for the target SNR can affect the quality of the fits, increasing it to higher values results in bigger bins and, consequently, in a loss of spatial information, especially in the outskirts of a galaxy. After fine-tuning based on the characteristics of our dataset (surface brightness and size of the dEs investigated in this study), we found a minimum SNR of $40$ to provide a better compromise between the spatial sampling of the galaxy outskirts and the accuracy of the measured kinematics. The spectrum assigned to each bin is the averaged spectrum of all spaxels within the defined bin. For a better accuracy of the fits, sky residuals particularly in the red part of the spectrum are masked. Later on, the stellar kinematics (i.e., rotational velocity and the velocity dispersion of the stellar component) of each galaxy were computed by fitting its binned spectra between 475 and 960 nm. 

For the fitting, we utilized the Extended MILES library (E-MILES) based on BaSTI isochrones \citep{2004ApJ...612..168P}. The E-MILES spectral library has moderately high resolution in the range of 1680-50000 \AA\, covering a relatively large range of ages (53 values, from 30 Myr to 14 Gyr) and metallicities (12 [M/H] values from -2.27 to +0.4). The spectral resolution of E-MILES is constant with a FWHM $\sim 2.51$ \AA \, \citep{2010Vazdekis, vazdekis2016}. In Appendix \ref{Instruments_resoloution} we discuss how the resolution of different spectral libraries affects our velocity measurements.

\begin{figure*}
\centering
\includegraphics[scale=0.78]{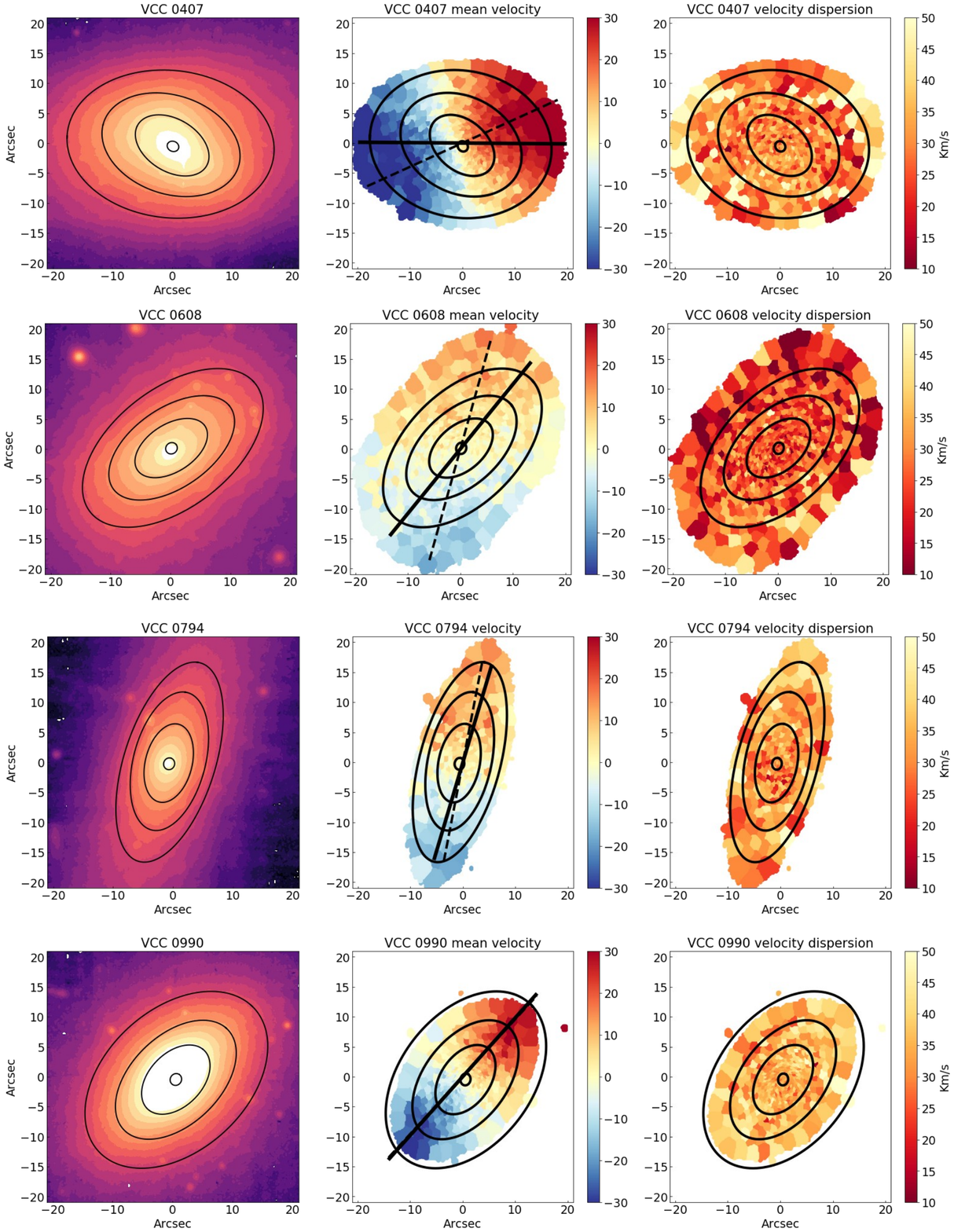}
\caption{Stacked MUSE images and stellar kinematic maps of our sample of dEs. In the left-hand panels, the stacked MUSE image of each galaxy is shown with black isophotes over-plotted. From the inside out, the isophotes indicate regions with a surface brightness of 20.59, 21.00, 21.85, and 22.52 mag/arcsec$^2$ (ABmag). In the middle panels, the line-of-sight velocity map of the stellar component of each dE is plotted. The kinematic position angle (PA$_{\rm kin}$) and photometric position angle of the major axis (PA$_{\rm phot}$) are traced with dashed and solid lines, respectively. In the right panels, the associated velocity dispersion map is shown. For a better comparison, the isophotes are drawn in all the panels. }
\label{set1}
\end{figure*}

\begin{figure*}
\centering
\includegraphics[scale=0.78]{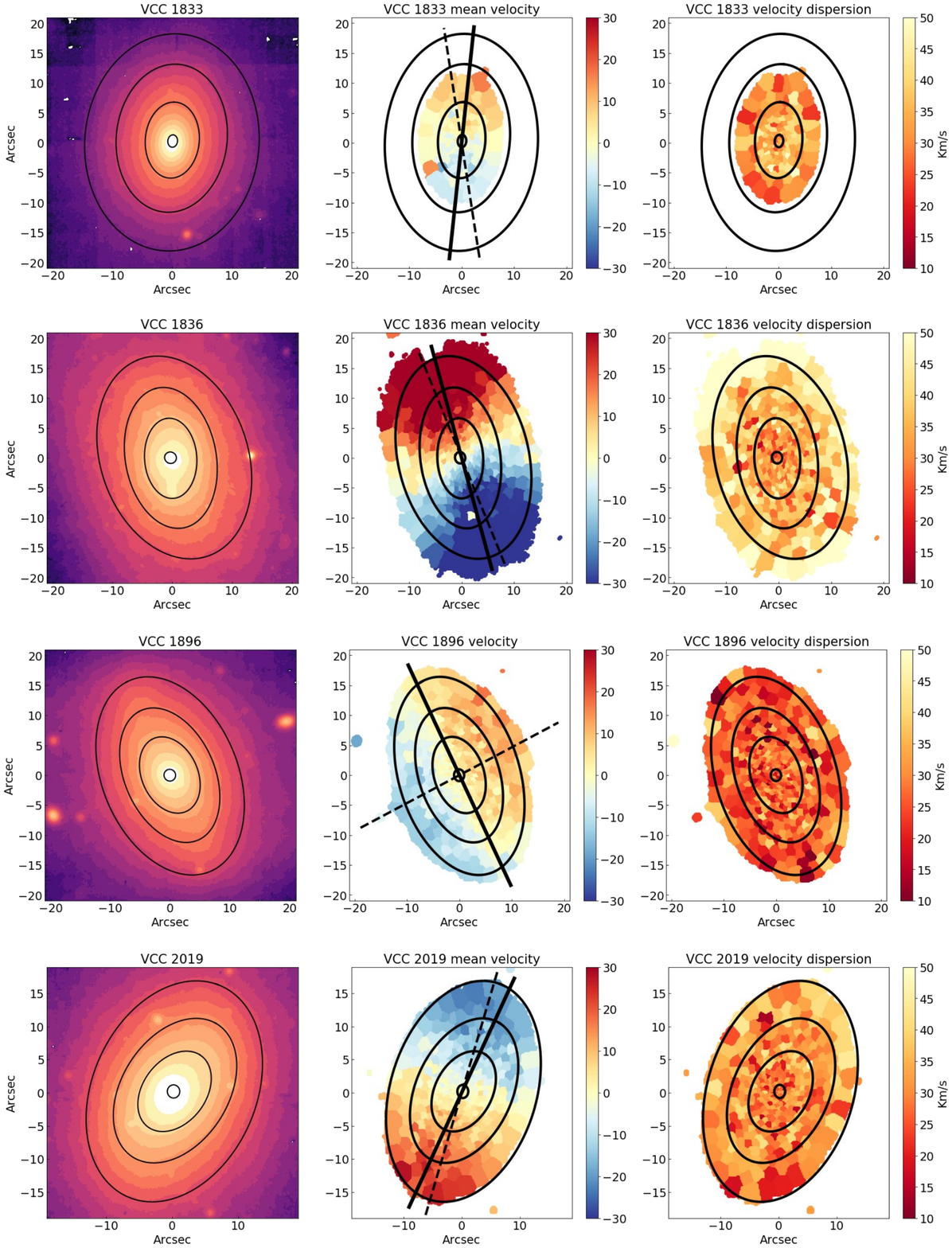}
\caption{Continued.}
\label{set2}
\end{figure*}

The E-MILES library was fitted to each bin's averaged spectrum through the penalized pixel-fitting algorithm (pPXF) introduced by \cite{Cappellari2004} and \cite{cappellari2017}. pPXF derives the line-of-sight velocity and velocity dispersion, parameterized via Gauss-Hermite moments, by using an approach of maximum penalized likelihood. In this study, the continuum slope was determined using additive Legendre polynomials available in pPXF with a degree of 6. As an example, in Fig. \ref{fitting_quality} we show the spectrum of a central bin in VCC0990. The plotted spectrum was chosen to show the quality of the data reduction and fits, as well as the degree of telluric lines contamination in the central regions of each dE in our data set. 

We obtained error estimates for each bin's averaged spectrum by using Monte-Carlo simulations. We ran a loop of 50 realizations for each spectrum. In each loop, we created a simulated spectrum from the original one by adding the fit residuals, randomly reshuffled among different wavelengths, to the original flux. The final reported error of each bin is the standard deviation of the velocity values obtained from the 50 realization loops. We present the error maps of velocity and velocity dispersion for each dE in Appendix \ref{Error_maps}.

\begin{figure*}
\centering
\includegraphics[scale=0.4]{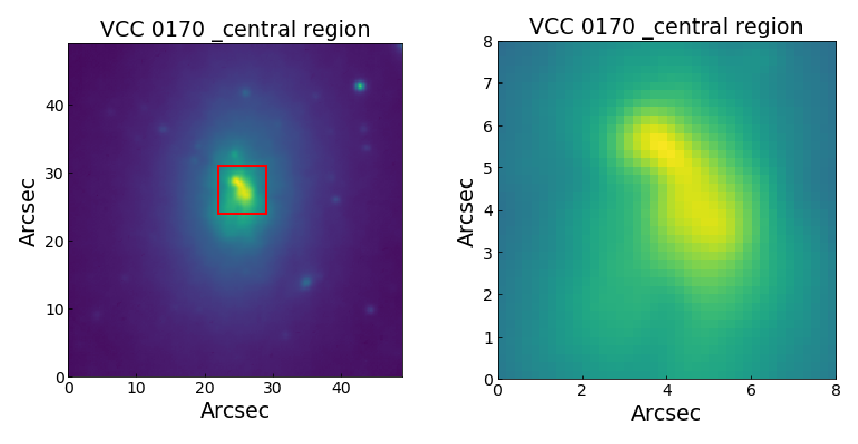}
\includegraphics[scale=0.3]{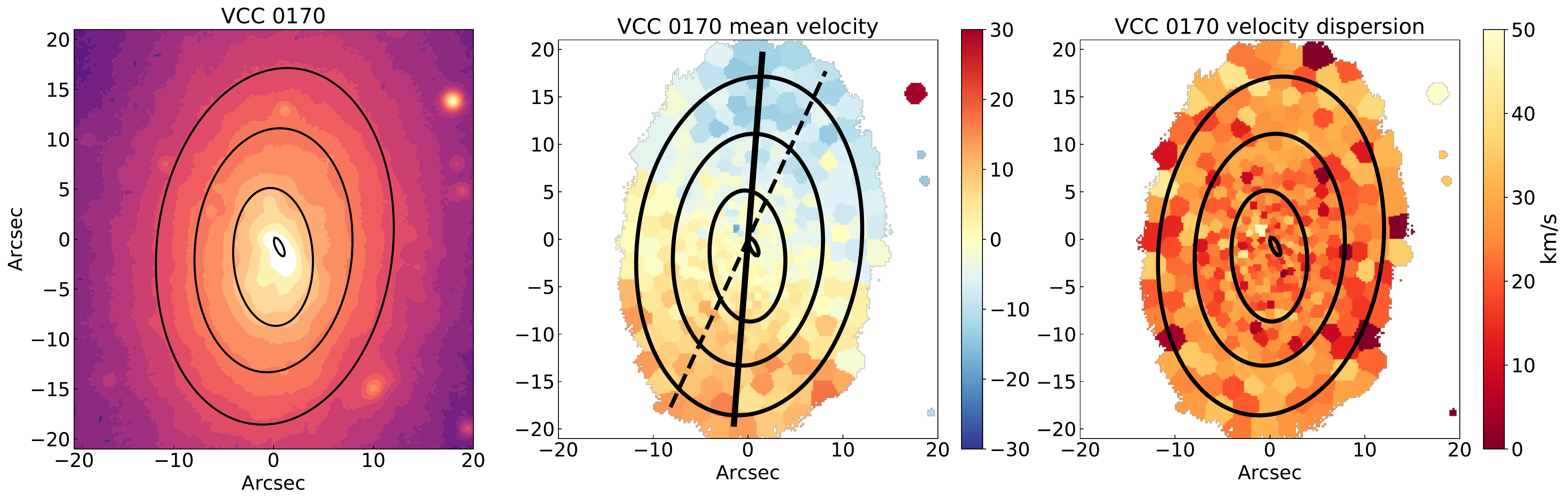}
    \caption{VCC0170 and its central gas component. \textit{Top panel}: On the left side, the MUSE stacked image is presented where the central region is marked with a red box, $8 \times 8$ arcsec$^{2}$ in size. In the right panel, a zoomed-in image of this particular region is provided, where the irregular shape of the core can easily be recognized. \textit{Bottom panels}: The kinematic maps of the stellar component in VCC0170. The left panel shows the MUSE stacked image while the rotation velocity and velocity dispersion of stellar component are plotted in the middle and right panels, respectively. Isophotes, kinematic and photometric position angles are also over-plotted as in Fig. \ref{set1}.}
\label{VCC0170Kin}
\end{figure*}

\begin{figure}
\centering
\includegraphics[scale=0.7]{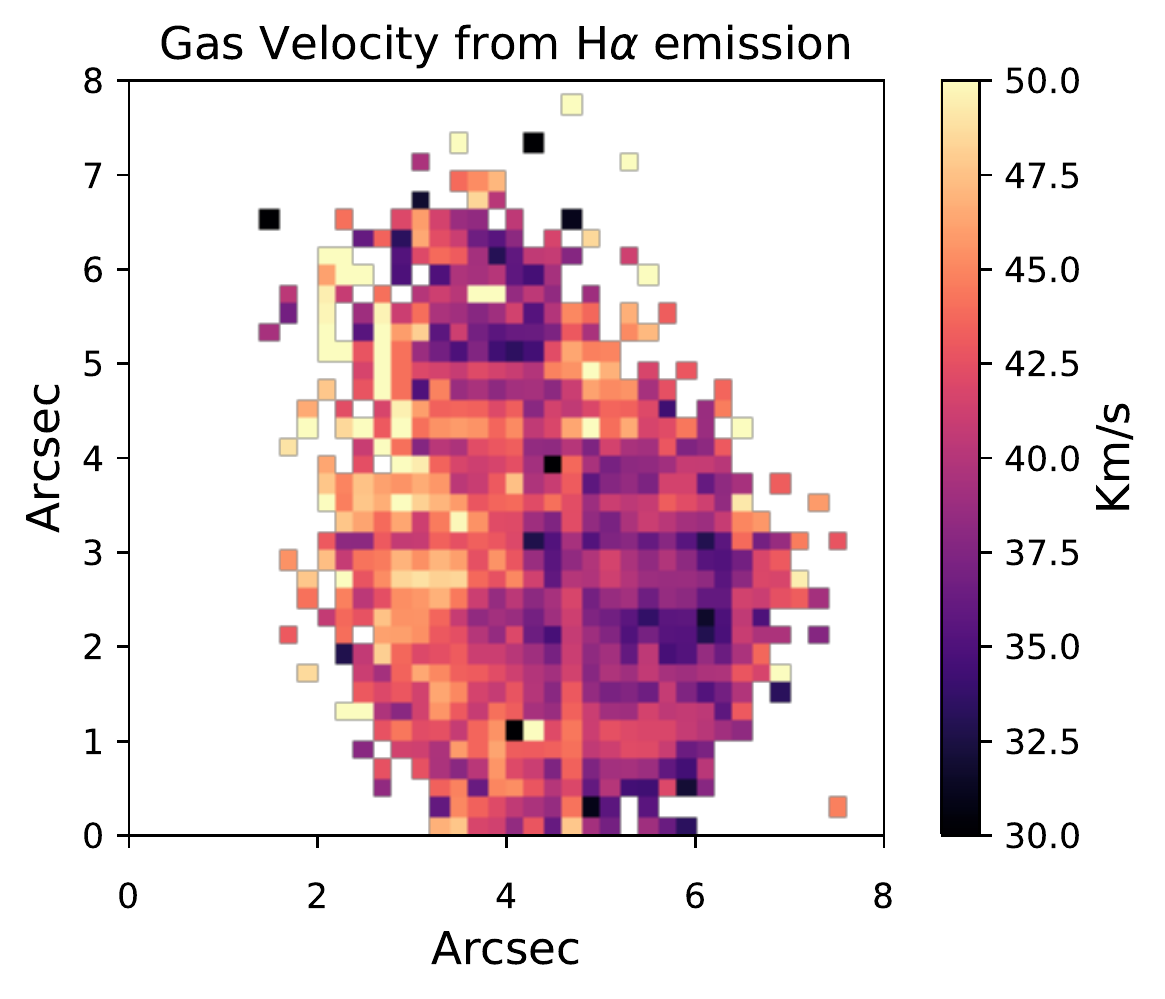}
    \caption{The gas velocity map from the H$\alpha$ emission line observed in the central part of VCC0170. This region is marked in the top right panel of Fig. \ref{VCC0170Kin}. The gas velocity here is relative to the stellar velocity.}
\label{VCC0170Kin_gas}
\end{figure}
\section{Results}\label{Results}
\subsection{Stellar kinematics}

For all of the targeted dEs, except VCC0170, the maps of stellar velocity and velocity dispersion along the line of sight are shown in Fig. \ref{set1} and \ref{set2}. Here, the left panels show the stacked data-cubes, obtained by averaging the flux across the full spectral range. For each image, isophotes (over-plotted in solid black lines) were measured by using the Ellipse method by \cite{1987Jedrzejewski}. The middle panels present the stellar velocity map of each dE, corrected for the galaxy's systemic radial velocity as measured by averaging the line-of-sight velocity of its bins within 0.1 $R_{\rm e}$ (to avoid offsets due to rotation of the galaxy in the outer regions). 
This radial velocity is reported in Table \ref{kin_result} and is in agreement with what was measured in SDSS DR15 \citep{2019Aguado}. For each dE, the maximum rotation velocity and velocity dispersion at 0.5 $R_{\rm e}$ are reported in Table \ref{kin_result}.

We find that the stellar component in our sample of dEs shows rotation with $V_{\rm max}$ ranging from $\approx$ 5 km\,s$^{-1}$ in VCC 1833 to $\approx$ 37 km\,s$^{-1}$ in VCC1836. 

\cite{2006Lisker} reported the likely presence of an inclined disk within VCC 0990 and VCC 0407. These objects show a relatively higher rotation with $V \geq 30$ km\,s$^{-1}$, consistent with the value obtained with long-slit spectroscopy by \cite{2015Toloba}.  

The velocity dispersion map of each dE is shown in the right column of Fig. \ref{set1} and \ref{set2}. As shown in the maps, the velocity dispersion profile of our sample of dEs has a shallow gradient with $ 20 \leq \sigma \leq 35$ km\,s$^{-1}$ within the field of view. A velocity dispersion gradient, although not statistically significant, is observed in VCC 1836, rising from 27.5 $\pm$ 3 km\,s$^{-1}$ in the central regions to more than 40.0 $\pm$ 18.0 \, km\,s$^{-1}$ in the outskirts of the galaxy.  

The stellar kinematics of VCC0170 was measured following the same steps as for the other dEs in our sample after masking the gas emission lines in its central region (Section \ref{VCC0170}). The velocity map of VCC 0170 shows rotation with $V_{\rm max} \approx \rm 18 $ km\,s$^{-1}$ in the outskirts of the galaxy. Like other members of our sample, VCC0170 has a flat velocity dispersion as a function of radius. The stacked image from the MUSE cube, the velocity map, and velocity dispersion map of VCC0170 are presented in the lower panels of Fig. \ref{VCC0170Kin}.

The position angle of the photometric major axis ($PA_{\rm phot}$) is overplotted with a solid black line on the velocity map of each galaxy. For each of our dEs we retrieved the $PA_{\rm phot}$ values measured in the g, r, and i bands from SDSS DR15 \citep{2015Alam}. They were obtained by fitting an exponential profile to the galaxy's observed surface brightness profile. For each galaxy we averaged the available PA values and computed the corresponding standard deviation. They are listed in Table \ref{kin_result}. On the same maps, the kinematic position angle ($PA_{\rm kin}$) of each galaxy is shown with a black dashed line. We measured the $PA_{\rm kin}$ using the kinematic maps of each dE following the method of \citet{2006MNRAS.366..787K}. To better quantify any offset, we computed the kinematic misalignment angle $\Psi$ defined by \cite{1991Franx}:

\begin{equation}
\sin\Psi = \mid sin(PA_{\rm phot} - PA_{\rm kin})\mid
\end{equation}

\noindent The resulting values are listed in Table \ref{kin_result}. 

VCC1896 has the highest misalignment between its photometric and kinematic position angles ($\Psi$ = 82.7 $\pm$ 3.1) and shows prolate-like rotation, with $V_{\rm max}$ $\approx$ 9.3 kms$^{-1}$ and $\sigma$ $\approx$ 23.0 kms$^{-1}$ consistent with what was measured by \cite{2015Penny}.

VCC0608 is another member of our sample that shows high kinematic misalignment ($\Psi$ = 42.8 $\pm$ 12.5). \cite{2006Lisker} reported a possible disk substructure within this galaxy. VCC0608 also exhibits an overall boxy shape in the image constructed from its MUSE data cube. We will discuss these two cases in more detail in Section \ref{rotation_0608_1896}.

\begin{table*}

\caption{\label{kin_result} Kinematic parameters of our nine dEs}
\centering
\begin{tabular}{c c c c c c c c c}
\hline

Object      &   $V_{\rm rad}$   & $V_{\rm max}$  &   $\sigma_{\rm 0.5R_{\rm e}}$  & $PA_{\rm ph}$  & $PA_{\rm kin}$  &  $\Psi$ & $\lambda_{Re}$ & $\lambda_{0.5Re}$\\
            &  [km\,s$^{-1}$] & [km\,s$^{-1}$]  & [km\,s$^{-1}$]   &   [deg]      & [deg]         & [deg] &  & \\
\hline
\hline
VCC 0170 &  1403.4  & 11.0 $\pm$ 15.0 &  24.5 $\pm$ 20  & 175.41 $\pm$ 0.51 &  155.2 $\pm$ 15.5  &  20.2 $\pm$ 15.5 & 0.45$^{\star}$     & 0.39\\
VCC 0407 &  1881.5  & 30.6 $\pm$  9.0 & 30.9 $\pm$ 18.0 & 89.56 $\pm$ 1.94 &  111.1 $\pm$ 3.1   &  20.4 $\pm$ 3.6& 0.67     & 0.40\\
VCC 0608 &  1807.7  & 7.8 $\pm$   14.0  & 21.6 $\pm$ 19.0 & 137.1 $\pm$ 1.77 &  180.0 $\pm$ 12.4  &  42.9 $\pm$ 12.5& 0.38     & 0.21\\
VCC 0794 &  1669.0  & 10.2 $\pm$  6.0  & 30.3 $\pm$ 20.0 & 161.7 $\pm$ 0.42 &  167.6 $\pm$ 3.1   &  5.9 $\pm$ 3.1 & 0.48$^{\star}$    & 0.27\\
VCC 0990 &   1715.1  & 10.5 $\pm$  4.0  & 31.5 $\pm$  6.0 & 134.5 $\pm$ 0.24 &  133.8 $\pm$ 4.6   &  0.5  $\pm$4.6 & 0.27   & 0.24\\
VCC 1833 &   1711.4  & 4.9 $\pm$   2.0  & 32.6 $\pm$  4.0 & 172.8 $\pm$ 0.38 &  189.6 $\pm$ 5.0   &  16.7 $\pm$ 5.0 & 0.15     & 0.11\\
VCC 1836 &  1985.2  & 36.9 $\pm$  7.0  & 34.6 $\pm$ 17.0 & 16.97 $\pm$ 0.18 &  24.8  $\pm$ 3.1   &  7.8  $\pm$ 3.1& 0.55$^{\star}$    & 0.53\\
VCC 1896 &   1872.5  & 9.3 $\pm$   7.0  & 23.0 $\pm$ 17.0 & 27.5 $\pm$ 0.58 &  110.2 $\pm$ 3.1   &  82.7 $\pm$ 3.1& 0.22   & 0.23\\
VCC 2019 &   1822.1  & 17.3 $\pm$  6.0  & 30.9 $\pm$ 15.0 & 153.2 $\pm$ 0.96 &  160.6 $\pm$ 9.3   &  7.2  $\pm$0.3 & 0.73   & 0.30\\
\hline
\end{tabular}\\

{The columns contain the name of the targets, the mean systemic line-of-sight velocity within 0.1 $R_{\rm e}$, maximum velocity ($V_{\rm max}$), and velocity dispersion ($\sigma$) at 0.5 $R_{\rm e}$, the position angle of the photometric major axis ($PA_{\rm ph}$) from \cite{2015Alam}, the kinematic position angle ($PA_{\rm kin}$), the kinematic misalignment angle ($\Psi$), the specific angular momentum at 1 $R_{\rm e}$ ($\lambda_{Re}$) and at 0.5 $R_{\rm e}$ ($\lambda_{0.5Re}$). Please note that values marked with a star are measured through extrapolation of the $\lambda_{R}$ profile.}\\

\end{table*}

Several studies of the Virgo cluster report the existence of so called kinematically decoupled cores (KDC) in early-type dwarf galaxies \citep[e.g., see][]{2003Geha, 2011Koleva, 2013Rys, 2014TolobaII, 2015ApJ...804...70G}. Observations of \cite{2014Toloba} show that about six percent of dEs in the Virgo cluster host a KDC. However, we do not detect any KDC in our sample of nine Virgo dEs. Although our sample is not statistically significant, thus not representative, the zero detection in this study is consistent with the results of \cite{2014Toloba}.

\subsection{Gas emission lines in VCC 0170}
\label{VCC0170}

Except for VCC0170 and VCC1836, our sample of dEs lacks nebular emission lines. In the case of VCC1836, the observed emission lines are produced by a background galaxy at redshift $z$ = 0.552 (see Appendix \ref{BB1_BB2} for a complete analysis).

The detection of nebular emission lines in the central part of VCC0170 is consistent with its blue colour, which is likely due to the presence of young massive stars. While this galaxy was originally classified as a late-type galaxy by \cite{2005Gavazzi}, it was recognized as a dE member of Virgo by \cite{2006Lisker}, who also pointed out an irregularly-shaped distribution of the blue colour in the central part of VCC0170. HI emission also is observed within the central part of this galaxy, indicating a $\frac{M_{HI}}{M_{bar}}$ ratio of less than one percent \citep{2005Gavazzi, 2006Lisker, 2006LiskerII}. Here $M_{\rm bar}$ denotes the baryonic mass.

As shown in the top right panel of Fig. \ref{VCC0170Kin}, the irregular shape of VCC0170's core is also visible in the MUSE stacked image. This is the region marked with a red box in the top left panel of Fig. \ref{VCC0170Kin}, which covers an area of 8$\times$8 arcsec$^{2}$ (0.40 kpc$^{2}$ at an assumed distance of 16.5 Mpc). A complete list of detected emission lines and their observed fluxes are reported in Appendix \ref{VCC0170_emissionlines}. Using the "Baldwin, Phillips, and Terlevich" (BPT) diagram \citep{1981Baldwin}, we confirm that the nebular emission from this irregularly shaped region is due to star formation. Furthermore, we have measured the gas kinematics for VCC0170 inside the marked region in Fig. \ref{VCC0170Kin} by fitting a Gaussian function to the H$\alpha$ line for each spaxel with SNR $\geqslant$ 3. We chose this particular emission line since it has a higher SNR in comparison to other available lines (see Fig. \ref{VCC0170Emission}). The resulting gas velocity map is shown in Fig. \ref{VCC0170Kin_gas}. We see that the gas component has a systematic offset in velocity ($\approx$ 40 $\pm$ \,8.0 km\,s$^{-1}$) with respect to the underlying stellar component.

\begin{figure*}
\centering
\includegraphics[scale=1.1]{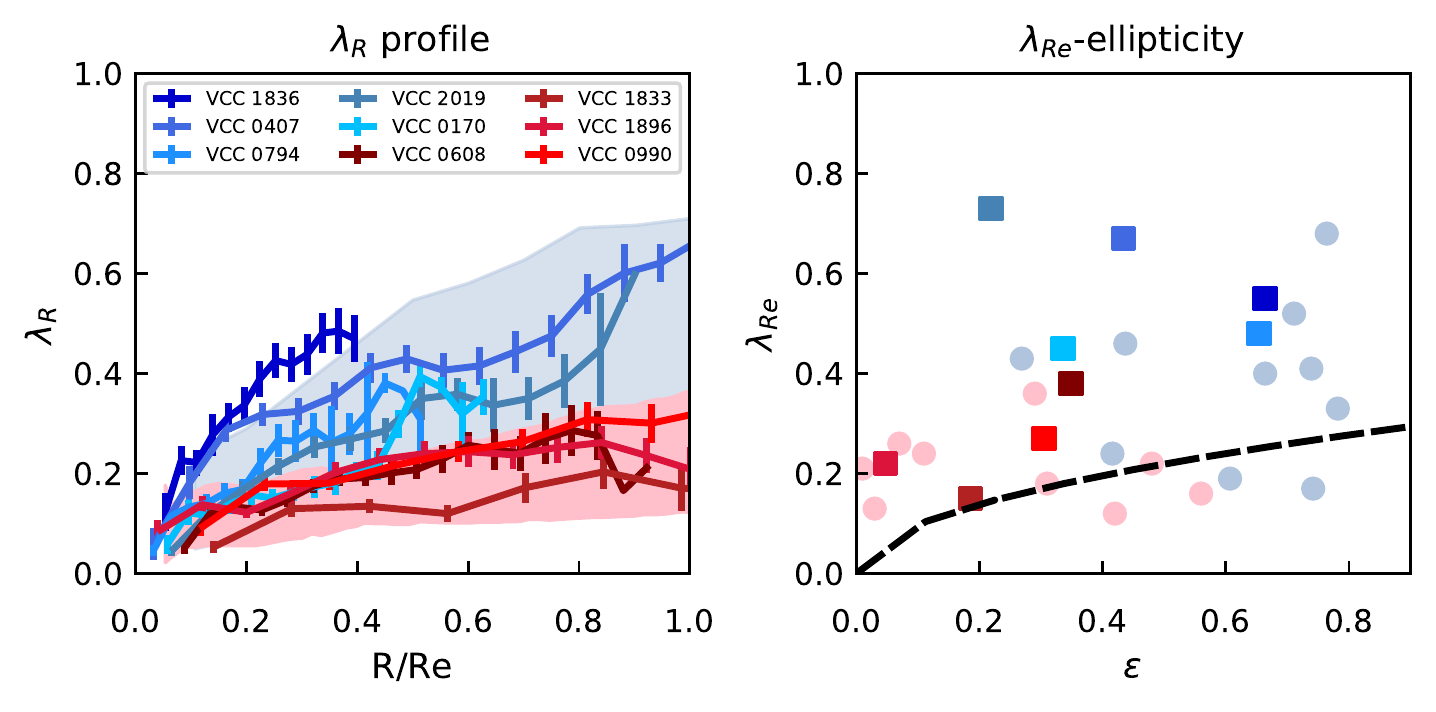}
\caption{\textit{Left panel:} The specific angular momentum profiles. The red-shaded area shows the minimum-to-maximum range of the $\lambda_{R}$ profiles of dEs in the sample. The blue-shaded region represents the minimum-to-maximum range of the $\lambda_{R}$ profiles of the CALIFA field galaxies. dEs in our sample with low $\lambda_{R}$ profiles similar to the R14 sample are traced with lines of different shades of red. Profiles of those dEs overlapping with the CALIFA galaxies are traced with lines of different shades of blue. \textit{Right panel:} $\lambda_{Re}$ -$\epsilon$ diagram distribution for our sample of dEs (colour-coded squares), dEs in the R14 sample (red circles) and ten CALIFA field galaxies from \citep{2019FalconBarroso} (blue circles). The dashed black line corresponds to the separating threshold between slow- and fast-rotators ($0.31 \times \sqrt{\epsilon }$) introduced by \citep{2011Emsellem}. }
\label{lambdaR}
\end{figure*} 

\subsection{Specific Angular Momentum}
\label{lambda_R}

To parameterize the rotation in our sample of dEs and in order to compare it with other dEs in the Virgo cluster and the CALIFA field galaxies, we derived the specific angular momentum ($\lambda_{R}$) profile of each dE, following \cite{2007Emsellem}:

\begin{equation}
\lambda_{R} = \frac{\sum_{i=1}^{N} F_{i}R_{i}|V_{i}|}{\sum_{i=1}^{N}F_{i}R_{i} \sqrt{V_{i}^{2} + \sigma_{i}^{2}}}
\end{equation}

\noindent where $R_{i},F_{i}, V_{i}$ and $\sigma_{i}$ are galactocentric distance, average flux, rotational velocity, and velocity dispersion of each bin, respectively. $R_{i}$ is the distance of each bin to the photometric center of the galaxy. The photometric center is measured using the Multi-Gaussian Expansion (MGE) fitting algorithm introduced by \cite{1994Emsellem} and adapted by \cite{2002Cappellari}.

The resulting $\lambda_{R}$
profiles are shown in the left panel of Fig. \ref{lambdaR} with different shades of blue and red. For a better comparison and in the same panel, the range of $\lambda_{R}$ profiles of dEs in the R14 sample is also shown by a red-shaded region. Similarly, the range of $\lambda_{R}$ profiles of the CALIFA field galaxies is shown by a blue-shaded region. Their $\lambda_{R}$ profiles were measured by \cite{2019FalconBarroso}.

In the left panel of Fig. \ref{lambdaR}, five dEs in our sample,
namely VCC0170, VCC0407, VCC0794, VCC1836, and VCC2019, show $\lambda_{R}$ profiles similar to those of the CALIFA field galaxies. We trace their $\lambda_{R}$ profiles with different shades of blue.
The remaining four dEs in our sample, namely VCC0608, VCC0990, VCC1833, and VCC1896,
exhibit $\lambda_{R}$ profiles comparable with the R14 sample (the red region), hence they
are traced with different shades of red.

Although the dEs investigated in this study have quite similar stellar masses, the large spread in the values and slopes of their $\lambda_{R}$
profiles is noticeable. VCC0170, VCC0407, and VCC2019 are characterized
by steeper profiles that rise from $\lambda_{R}$ = 0.06 at the centre to  
$\lambda_{R}$ =
0.6 at about 1 $R_{\rm e}$. The $\lambda_{R}$ profile of VCC1836 shows a truncation at about 0.40 $R_{\rm e}$. This galaxy is the most extended member of our sample and the MUSE field of view is too small for sampling it to 1 $R_{\rm e}$ . The same explanation applies to the $\lambda_{R}$
profiles of VCC0170 and VCC0794, which only reach 0.60 $R_{\rm e}$ and 0.50 $R_{\rm e}$,  
respectively.

In the right panel of Fig. \ref{lambdaR}, we plot the $\lambda_{Re}$-$\epsilon$ diagram ($\lambda_{Re}$ vs. ellipticity) for the same set of galaxies as shown in the left panel. Here $\lambda_{Re}$ is the value of the specific angular momentum at one effective radius. In this diagram, the dEs investigated in this study are denoted with squares and are colour-coded as in the left panel of the same figure. dEs in the R14 sample and the CALIFA field galaxies are denoted with pink and blue circles, respectively. As mentioned earlier, three dEs in our sample are too extended for the MUSE field of view to be mapped up to 1 $R_{\rm e}$. For these three galaxies, the $\lambda_{Re}$ value at 1 $R_{\rm e}$ is measured through extrapolation of their $\lambda_{R}$ profiles. The obtained values of $\lambda_{R}$ at 1 $R_{\rm e}$ and 0.5 $R_{\rm e}$ are reported in Table \ref{kin_result}. The black dashed line in Fig. \ref{lambdaR} corresponds to ($0.31 \times  \sqrt{\epsilon }$), which is the threshold separating slow from fast-rotating galaxies, according to \cite{2011Emsellem}. The $\lambda_{\rm Re}$-$\epsilon$ diagram indicates that, at a fixed stellar mass, on average the CALIFA field galaxies show higher values of $\lambda_{R}$ and ellipticity in comparison to dEs in Virgo. The low $\lambda_{R}$ dEs in our sample are mostly distributed close to the boundary separating the fast- from the slow-rotating galaxies. As mentioned in Section \ref{observation}, our sample of dEs are located in the outskirts of the Virgo cluster and according to the right panel of Fig. \ref{lambdaR}, they are fast-rotating systems. This is consistent with the results of \cite{2014Bosellib} who showed that fast-rotating galaxies are mostly observed in the outskirts of the Virgo cluster.

\section{Discussion} \label{discussion}

\subsection{Rotation in dEs}
\label{rotation}

The kinematic maps of our sample of dEs show different degrees of rotation (Fig. \ref{set1} and \ref{set2}). This result is consistent with other studies of Virgo dEs in the literature. \citet[hereafter T11]{2011Toloba} investigated a sample of 21 dEs in Virgo with $-15\geq M_{r}> -18 $ for which they measured a similar range of rotational velocities, from 0 to 50 km\,s$^{-1}$. The velocity dispersion of the T11 sample also varies between 30 to 50 km\,s$^{-1}$. Meanwhile, by focusing on a smaller sample of Virgo dEs with a similar range of M$_{r}$, R14 investigated 12 dEs and showed that their rotational velocity varies from 0 to 40 km\,s$^{-1}$. The velocity dispersion maps of dEs in that study also show a flat radial gradient, with values ranging from 40 to 60 km\,s$^{-1}$. The values of rotational velocity for the dEs investigated in this study are in the same range as in T11 and R14 but skewed toward higher values, especially at higher galactocentric distances.

To compare the physical properties of dEs with the aim of investigating the role of the environment in their transformation, one needs to consider their infall time (i.e., how long a dE was exposed to environmental effects). That is mainly due to the fact that environmentally induced mechanisms such as tidal interactions and ram pressure stripping occur over different but not short
time scales \citep{2006Boselli, Tinker_2010, 2015Bialas}. According to \cite{2006Boselli}, ram pressure stripping occurs with a typical timescale of less than 1 Gyr and tidal interactions happen on a timescale of $\sim$ 2 Gyr, which corresponds to the crossing time of the Virgo cluster. Average values of infall time can be derived using a galaxy's position in the projected phase-space diagram of their host halo \citep{2017Rhee, 2019Pasquali, 2019Smith}. Based on their position in the observer's phase-space (see the lower panel of Fig. \ref{Virgo_distribution}), we know that our sample dEs are in their initial phase of accretion to the Virgo cluster, i.e., they were accreted to Virgo around 2-3 Gyr ago \citep{2015Vijayaraghavan, lisker2018}. On the other hand, the position of the R14 and T15 dE sample in the observer's phase-space diagram reveals that around 60 per cent of them had an earlier average infall time than our sample of dEs. The difference in accretion time may explain the overall higher rotational velocity of our sample dEs, in comparison to the population investigated by R14. We notice here that, when inferring the average infall times from the projected phase-space diagram of galaxy clusters, \cite{2019Pasquali} consider accretion to the whole cluster and not infall onto its dominant structures or substructures. The latter case would possibly deliver more accurate infall times in the case of the Virgo cluster, which is noticeably structured. Thus, the comparison of infall time of the different dE samples considered in this paper should be taken with some caution.

Furthermore, results of \cite{2017A&A...597A..48F} for the CALIFA field galaxies show higher values of rotational velocity (from 20 to 100 km\,s$^{-1}$) and velocity dispersion (from 20 to 90 km\,s$^{-1}$). Our sample of dEs show lower degrees of rotation in comparison to the CALIFA field galaxies. As already discussed in Section \ref{introduction}, such differences between the kinematics of low-mass field galaxies and Virgo dEs can be expected due to the cluster's role in transforming its accreted members. Besides, prior to their infall to Virgo, our sample of dEs belonged to a rather massive group which models predict to have been as massive as $10^{13}$ M$_{\odot }$ \citep{lisker2018}. Therefore, the lower degrees of rotation observed in our sample dEs in comparison to the CALIFA field galaxies, may be understood in light of pre-processing in their parent group. This will be discussed with more detail in Section \ref{LAMBDAr_DISC}.

\subsection{Stellar rotation in VCC0608 and VCC1896} 
\label{rotation_0608_1896}
Among the dEs investigated in this study, VCC0608 and VCC1896 present a rather high offset between their photometric and kinematic position angles. These two dEs can not be classified as regular rotators. In this subsection we will discuss different possible explanations for the presence of such an irregular kinematic structure.

VCC1896 with $\Psi$ = 82.7$^{\circ}$ $\pm$ 3.1$^{\circ}$ has the highest offset between its photometric and kinematic axis. Taken at face value, VCC1896 would look like a dE with minor axis rotation. \cite{2006Lisker, 2009Lisker} reported the existence of two rather faint spiral arms for VCC1896 and a bar aligned with the outer photometric minor axis of this galaxy. \cite{2005MastropietroII} showed that, due to tidal interactions, asymmetric features and open spiral arms can be formed in those dEs which are located in the outskirts of cluster. In addition to that and as shown by \cite{2019kwak}, the cluster tidal field can trigger the formation of weak spiral arms in disk dEs. These results might explain the existence of the faint spiral arms in VCC1896. Due to their low surface brightness, the faint spiral arms of VCC1896 are not detected in our MUSE data cube; therefore, they are not present in the kinematic maps of this dE. We only detected the bright bar of this dE in its MUSE cube. While the presence of the bar explains the high offset between kinematic and photometric PA, it may also be a morphological feature relevant to the evolution of VCC1896.

The bar in the central region of VCC1896 can either be a pristine structure that formed and developed through standard processes or a tidal feature that formed as a result of interactions with the environment \citep[e.g., see][]{2005MastropietroII}. Tidal interactions are known as one of the most dominant mechanisms that can induce instabilities in the stellar disk of galaxies and form bars, particularly in central regions of galaxy clusters and groups \citep{ 2016ApJ...826..227L}. Moreover, \cite{2019kwak} showed that tidal effects due to galaxy-galaxy interactions, which are more dominant in the central parts of a cluster, can be efficient in causing instabilities to bar formation in dwarf galaxies in a Virgo-like halo. As for Virgo, observations of \cite{2012Janz} also confirm that dEs with bars are mostly observed in the central parts of the cluster. Also, simulations have shown that bar instabilities can be triggered in Local Group dwarf galaxies after their first pericenter passage \citep{2016Gajda}. However, based on \cite{2019Pasquali}, the probability of having already experienced a first pericenter passage for galaxies with a similar average infall time as VCC1896 is not particularly high. Thus, the  presence  of  a  bar  structure in VCC1896 may be interpreted in the light of pre-processing in its parent group, and less likely in Virgo. 

\begin{figure}
\centering
\includegraphics[scale=0.6]{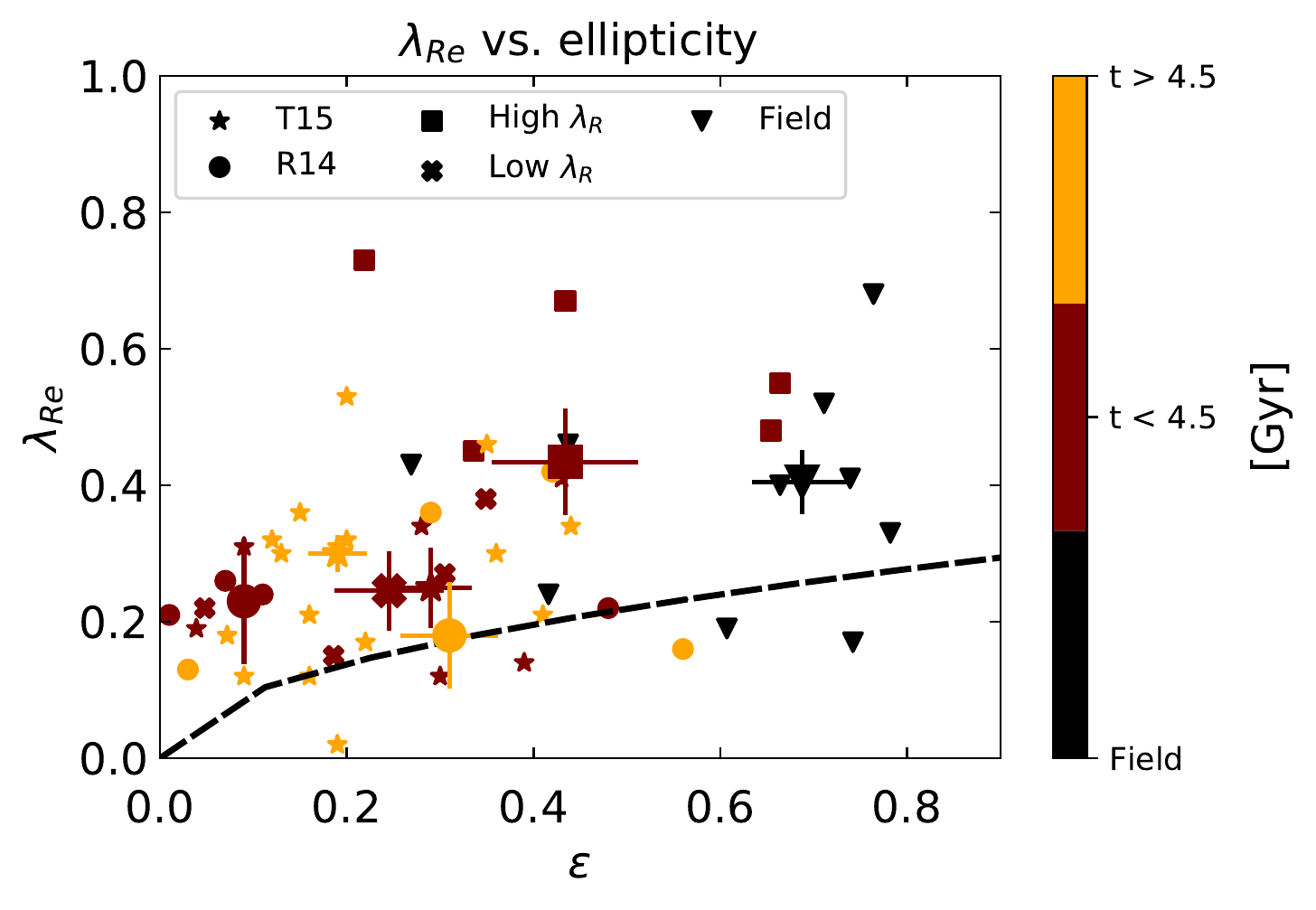}
    \caption{The $\lambda_{Re}$-$\epsilon$ diagram colour coded based on the average infall time of each galaxy. The CALIFA field galaxies are denoted with black triangles. Our sample of dEs with high and low values of $\lambda_{Re}$ are traced with squares and crosses, respectively. The R14 sample dEs are marked with circles. The T15 sample dEs are marked with stars. The median ellipticity and $\lambda_{Re}$ of each group of galaxies are also denoted with their corresponding symbol, but bigger, and error bars. The dashed black line corresponds to the separating threshold between slow- and fast-rotators ($0.31 \times \sqrt{\epsilon }$) introduced by \citep{2011Emsellem}.}
\label{average_profiles}
\end{figure}

VCC0608 with $\Psi$ = 42.9$^{\circ}$ $\pm$ 12.5$^{\circ}$ is the second dE in our sample showing significant misalignment between its kinematic and photometric minor axis. The observed kinematic misalignment can be the trace of past or on-going tidal interactions with the massive halo of Messier 100 (projected distance to M100 $\sim$ $28.7$ kpc at an assumed Virgo distance of 16.5 Mpc). VCC0608 is a boxy-shaped galaxy as its isophotes on the MUSE stacked image show A$_{4}$/a $\approx$ -0.02 $\pm$ 0.003 at 1 $R_{\rm e}$. 
Therefore, VCC0608 shares similarities with LEDA 074886 \citep{2012graham}, which is a boxy-shaped dwarf galaxy with M $\approx$ $10^{9}$ M$_{\odot}$ in the halo of NGC 1407. Following the same argument as in \cite{2012graham}, the boxy shape of VCC0608 and its high kinematic misalignment angle may indicate that this galaxy is the remnant of a past nearly edge-on merger of two disk dwarf galaxies \cite[see e.g.,][]{2006Naab}. 
In this case, the inner disk of the galaxy is believed to have formed from gas driven inward during the merger event \citep{2012graham}. In that case, a younger stellar population is expected to be observed in the central region of VCC0608. This scenario will be tested in a forthcoming paper through the analysis of the stellar populations of our sample of dEs.

\subsection{Group vs. cluster environment}
\label{LAMBDAr_DISC}

Based on the kinematic analysis performed in this study, we found that the specific angular momentum ($\lambda_{R}$) profile of our sample of dEs is intermediate between that of dEs in Virgo and CALIFA low-mass galaxies in the field.  

The angular momentum is often described as an essential metric for understanding the evolution of low-mass galaxies as it may reflect conditions of their early environments where these objects formed or spent most of their lives \citep[see e.g.,][]{2015Shi}. The angular momentum is typically parametrized in terms of the projected stellar specific angular momentum along the line of sight ($\lambda_{R}$). This parameter is weighted according to the luminosity, thus it is considered to be a relevant proxy for the real stellar specific angular momentum of a galaxy \citep{2007Emsellem, 2009Jesseit, 2011Emsellem}. Studies have revealed that, through the lifetime of a galaxy, mechanisms such as gas accretion and star formation can effectively increase the angular momentum of the entire structure. For instance, \cite{2019Zoldan} showed that at a fixed stellar mass, gas-rich systems tend to have higher degrees of rotation. Meanwhile, other studies show that mergers and tidal heating of the disk can decrease the net rotation of a galaxy and thus its angular momentum \citep[for example, see][]{2001vandenBosch,2011Brook,2014Naab,2017Penoyre,2018Greene}. Tidal interactions occur frequently in galaxy groups or central parts of a galaxy cluster, mostly due to interactions with the halo's tidal field and  encounters with other cluster members. Tidal interactions are known to be one of the effects responsible for the evolution of $\lambda_{R}$ as well as for the morphological transformation of galaxies in clusters and groups  \cite[e.g.,][]{2006Lisker, 2006Boselli, 2007Lisker, 2008Michielsen, 2009Toloba, 2009Lisker, 2009Kormendy, 2010Geha, 2011Toloba,2012Janz, 2012Toloba, 2014Rys}. 
On the other hand, ram pressure stripping is expected to have a marginal effect on altering the $\lambda_{R}$ profile of a galaxy \citep{2016Yozin}. 
Following the scenario suggested by R14, cluster low-mass galaxies with earlier infall time (i.e., with more pericenter passages) are expected to experience a more significant tidally-driven change of $\lambda_{R}$. Accordingly, newly accreted low-mass galaxies are expected to exhibit internal kinematics intermediate between that of low-mass field galaxies and  "more ancient infallers" in cluster. Such expected intermediate kinematics is shown in the left panel of Fig. \ref{lambdaR}. In this plot, our sample dEs, with an average infall time of 2-3 Gyr to Virgo, distribute between the CALIFA low-mass galaxies in the field and the Virgo dEs in the R14 sample. The latter sample has been exposed to environmental effects for longer time ($\sim$ 1-2 Gyr), according to the lower panel of Fig. \ref{Virgo_distribution}.

Along with $\lambda_{R}$, the ellipticity of a satellite galaxy is also expected to decrease with infall time on the ground of tidally-driven effects that can thicken the stellar disk of a system. For instance, \cite{2015MNRAS.451..827D} show that in a rich cluster, the fraction of spheroidal early-type galaxies increases toward lower cluster-centric distances (but see: \cite{2014Weijmans}).
In Fig. \ref{average_profiles} we represented the same $\lambda_{Re}$-$\epsilon$ diagram of Fig. \ref{lambdaR} where different groups of galaxies are now colour-coded according to their average infall time, as read from the bottom panel of Fig. \ref{Virgo_distribution}. dEs with high and low values of $\lambda_{Re}$ in our sample are marked with squares and crosses, respectively. The R14 dE sample is denoted by circles and, additionally, the T15 sample is indicated with stars. In Fig. \ref{average_profiles}, the CALIFA field galaxies (denoted with triangles) with an assigned average infall time of zero Gyr exhibit, on average, higher ellipticity than Virgo dEs. The average ellipticity and $\lambda_{Re}$ of each group are indicated by their corresponding symbol, but bigger, and error bars. A spread, similar to that of the $\lambda_{R}$ profiles, is also visible in the ellipticity of the nine dEs investigated in this study. The ellipticity of the low-$\lambda_{R}$ dEs in our sample is comparable to those dEs in the T15 sample of similar infall times. The ellipticity of low-$\lambda_{R}$ dEs in our sample is also comparable to those dEs in the R14 sample that have, on average, earlier infall times.

Given their similar stellar masses and comparatively short infall time to Virgo, the spread in the $\lambda_{R}$ profile of our sample of dEs is rather large (Fig. \ref{lambdaR}). One possible explanation is that the Virgo environment has had only marginal impact on the
evolution of the $\lambda_{R}$ profiles of our sample of dEs (if any) on the ground that their recent infall time decreases their probability of having already experienced a pericenter passage. In this regard, \cite{2019Pasquali} showed that near 40 per cent of satellite galaxies with an average infall time of 3.6 Gyr have not experienced their first pericenter passage yet. Environmental processes during pericenter passages are more efficient in altering $\lambda_{R}$ rather than at higher cluster-centric distances \citep{2012Villalobos}. 

According to the scenario proposed by \cite{lisker2018}, prior to their infall to Virgo, our sample of dEs were probably members of a rather massive galaxy group  ($10^{13}$ \(\rm M_\odot\)) where the probability of tidal interactions is high \citep{2009Tal} due to lower velocity dispersion of the halo \citep{2006Boselli}. In fact, \cite{2012Villalobos} have shown that in an equally massive halo, the $\lambda_{R}$ profile of a galaxy progressively flattens and decreases in value as the galaxy undergoes more and more pericenter passages. The latter is mainly by virtue of substantial dynamical disturbances that arise often after several orbital periods (i.e. $>$ 4 Gyr). In this regard, the diversity in ellipticity and $\lambda_{R}$ profiles of our sample of dEs may be a reflection of different infall times in their previous host halo. Hence, the four dEs with $\lambda_{R}$ profiles as low as those of the Virgo dEs in the R14 sample (in red in Fig. \ref{lambdaR}) may have been accreted earlier onto their previous host halo, and therefore evolved more strongly under their group's environmental mechanisms. They include two galaxies with footprints of tidal interactions in their kinematics (i.e., VCC1896 and VCC0608 in Section \ref{rotation_0608_1896}). Contrarily, the remaining five dEs with $\lambda_{R}$ profiles similar to those of the CALIFA field galaxies (blue profiles in Fig. \ref{lambdaR}) may have experienced partial transformation mainly due to their comparatively lower number of pericenter passages in this previous host halo, if any. In this picture, what we have measured for our sample of dEs may be considered as the result of pre-processing in a different environment than Virgo.

Stripping the gas content of galaxies is a known effect of high-density environments (i.e., massive galaxy groups and clusters) mainly due to ram pressure stripping \citep{2006Boselli, 2008Boselli}. The efficiency of such mechanism depends directly on the density of the inter-cluster medium, as well as on the speed and orbit of the accreted galaxy \citep[e.g.,][]{1972Gunn, 2001Vollmer,2020Steyrleithner}. Ram pressure stripping can be noticeably efficient for low-mass galaxies, mainly due to their shallow potential well. According to \cite{2008Boselli} and \cite{2014Bosellib}, low-mass systems, if gas rich, can become quiescent systems on rather short timescales in the Virgo cluster. Thus, both the previous parent group and the Virgo cluster could have been responsible for gas stripping in our sample of dEs \citep[for gas stripping in groups see][]{2017Brown}. Moreover, \cite{2016McPartland} and \cite{2019Ebeling} show that
 the efficiency of ram pressure stripping is highest when the velocity vectors of galaxies in a group being accreted onto a cluster are aligned, thus add up to the velocity vector of the infalling group itself.
 
 Ram pressure can perturb and make the gas content of galaxies unstable \citep{2008Kronberger, 2020Steyrleithner}. We observe these effects in VCC0170, whose gas component exhibits a velocity offset from the galaxy's stellar body, and nebular emission lines consistent with recent/on-going star formation (see Appendix \ref{VCC0170_emissionlines}) \citep{2014Roediger}. This finding allows us to envisage an alternative scenario. In our sample, those dEs with higher values of $\lambda_{R}$ could have been low-mass, late-type, star-forming galaxies in the previous host halo, which lost most of their gas reservoir after infall to Virgo, mostly via ram pressure stripping \citep{2009Toloba, 2014Bosellib}. As already explained in the first scenario, tidal interactions only develop during several orbital periods, comparable to the cluster dynamical time-scale. Therefore, our sampled dEs with high but comparable values of $\lambda_{R}$ and ellipticity to field galaxies have experienced little dynamical transformation thus far in Virgo. In this picture, the low-$\lambda_{R}$ dEs in our sample could have been accreted to Virgo as early-type galaxies that likely lost their gas content and were transformed dynamically in their previous environment. Moreover, tidal disturbances (such as in VCC0608 and VCC1896) indicate tidal encounters that occurred in the past, but possibly not in Virgo, due to their rather recent infall time. We then suggest that the low-$\lambda_{R}$ dEs in our sample have likely been processed by their previous group in both the first and second scenarios. This second scenario is more in line with what suggested by \cite{2009Toloba, 2011Toloba, 2015Toloba}. Further analysis on age, metallicity, and star formation history of this sample can provide us with a better understanding of their evolution before and
after accretion onto the Virgo cluster and thus evaluate this scenario. The results of such an analysis will be presented in a forthcoming paper.

Alternatively, the observed spread in the $\lambda_{R}$ profiles and ellipticity of our sample could be consistent with a recently accreted population of field dEs, with intrinsically different degrees of rotation and ellipticity. \cite{2017Janz} studied a sample of nine isolated quenched dEs in the Local Volume and reported diverse kinematics for this sample. 
In this regard, the four low-$\lambda_{R}$ dEs in our sample could originally have been primordial slow-rotating objects with a rather flat $\lambda_{R}$ profile even before entering a high-density environment. The latter is consistent with the results of \cite{2017Wheeler}, who show that in the absence of external tidal fields, isolated dwarf galaxies can be formed as dispersion-dominated systems. In this alternative interpretation, the dEs with higher values of $\lambda_{R}$ can intrinsically be different from their low-$\lambda_{R}$ counterparts in our sample, exhibiting kinematics due to relatively little or moderate processing thus far in the cluster. \cite{2012Geha} showed that the fraction of field dEs in the same stellar mass range of our sample and with no active star formation is less than 0.2 percent (corresponding to a total number of 4 galaxies). According to this study, we cannot firmly rule out the possibility that the low-$\lambda_{R}$ dEs in our sample were also field members accreted to Virgo. However, it would be a rare coincidence that these four field galaxies were accreted on to Virgo with a similar velocity as the accreted group and at the same time. We believe that constructing a more statistically rich sample of dEs in different environments (such as field, groups, and galaxy clusters) can provide us with a more complete picture of the role of environment in altering kinematics and other properties of low-mass galaxies.


\section{Conclusions} \label{conclusion}

In this study, we investigated the stellar kinematics of nine dEs belonging to a group that was accreted onto the Virgo cluster, along the observer's line of sight, about 2-3 Gyr ago \citep{lisker2018}. Their similar accretion time has also been confirmed by the position of these dEs in the phase-space diagram of \cite{2019Pasquali} and \cite{2019Smith}. Members in this sample provide a unique test bed to study the evolution of dEs in their initial phases of infall to Virgo. We obtained MUSE IFU data cubes for these nine dEs in order to investigate their kinematics, dynamics, and stellar populations. In this work, we present the kinematics of these nine dEs. We measured the line-of-sight velocity and velocity dispersion across the stellar body of each dE, using pPXF on the full spectral range delivered by MUSE. Later on, we used these values in order to compute the $\lambda_{R}$ profiles of our sample dEs with the purpose of comparing them with other similar measurements for different environments available from the literature. The main results of our analysis can be summarized as follows:

\begin{itemize}
    \item The stellar component of our sample of dEs shows a rather similar range of rotational velocities in comparison to other dEs in Virgo (from 5 to 40 km\,s$^{-1}$). The velocity dispersion of their stellar component varies between 20 to 35 km\,s$^{-1}$. 
    
    \item We investigated the kinematic misalignment angle of the dEs in our sample, and we report two cases of misaligned kinematics, namely VCC1896 and VCC0608. In the case of VCC1896, we show that the observed offset between the kinematic and photometric position angle is basically due to the presence of a bright bar that dominates the photometric measurements. We also find a high kinematics misalignment for the boxy-shaped VCC0608. Past or on-going tidal interactions with the massive halo of M100 or a past major merger, considering similarities between this particular dE and the galaxy LEDA 074886, may be possible causes of the observed high kinematic misalignment in this galaxy. 
    
    \item VCC0170 is the only dE in our sample that shows strong nebular emission lines in its very central region due to star formation, presumably triggered by recent/on-going gas stripping in Virgo. Further analysis of the star formation history and metallicity of this particular dE (to be presented in a forthcoming paper) may shed light on the origin of the asymmetrical structure in the center of VCC 0170. 
   
    \item The $\lambda_{R}$ profile and ellipticity of the dEs investigated in this work show, on average, intermediate values between the dEs studied by \cite{2014Rys}, which were on average accreted earlier onto Virgo, and CALIFA field galaxies of the same stellar mass (from \cite{2019FalconBarroso}). 
    
    \item Given their infall time to Virgo and similar stellar masses, the spread in the  $\lambda_{R}$ profile of our nine dEs can be interpreted as follows: the low-$\lambda_{R}$ dEs were likely transformed in and by their previous host group, prior to their infall onto Virgo, while the high-$\lambda_{R}$ dEs were only partially transformed in their previous host group. The latter sub-group of dEs may be experiencing ram pressure stripping in Virgo.
    
    \item In Appendix \ref{BB1_BB2} we report on the serendipitous discovery of two intermediate redshift galaxies, found in the MUSE data cubes of VCC0170 and VCC1836.

\end{itemize}

\begin{footnotesize}
\section*{Acknowledgments}

We thank the referee for useful suggestions that
improved the clarity of this paper. We acknowledge financial support from the European Union's Horizon 2020 research and innovation program under the Marie Sklodowska-Curie grant agreement no. 721463 to the SUNDIAL ITN network. We would like to thank Jakob Walcher, Jorge Iglesias, J.K Barrera-Ballesteros, D.J. Bomans, A. Miskolczi, and B. Biskup for providing environmental information for the CALIFA sample. B.B. acknowledges the support of the International Max Planck Research School (IMPRS) for Astronomy and Cosmic Physics at the University of Heidelberg. J. F-B  acknowledges support through the RAVET project by the grant AYA2016-77237-C3-1- P from the Spanish Ministry of Science, Innovation and Universities (MCIU) and through the IAC project TRACES which is partially supported through the state budget and the regional budget of the Consejer\'ia de Econom\'ia, Industria, Comercio y Conocimiento of the Canary Islands Autonomous Community. GvdV acknowledges funding from the European Research Council (ERC) under the European Union's Horizon 2020 research and innovation programme under grant agreement No 724857 (Consolidator Grant ArcheoDyn). 

\end{footnotesize}

\section{Data Availability}
This work is mainly based on observations collected at the European Southern Observatory under P98, ESO programmes 098.B-0619 and 0100.B-057. 

\bibliographystyle{mnras} 
\bibliography{Bidaran2020}

\newpage
\begin{appendix}
\section{CALIFA field galaxies}\label{Califa_galaxies_appendix}
The Calar Alto Legacy Integral Field Area (CALIFA) survey was designed to investigate the evolution of galaxies through cosmic time by a detailed spectroscopic study of $\sim$ 600 galaxies in the Local Universe \citep{2012A&A...538A...8S, 2014Walcher}. As our field control sample, we selected 10 low-mass field galaxies from the CALIFA survey. The selected CALIFA objects are star-forming galaxies of similar stellar mass to those observed in this study, following the criteria mentioned in Section \ref{lambda_R}. We summarize the properties of these 10 galaxies in Table \ref{CALIFA_TARGETS}.

\begin{table*}
\caption{\label{CALIFA_TARGETS} List of CALIFA targets}
\centering
\begin{tabular}{c c l l c c c  c c}
\hline
Object      & type    & $\alpha$ (J2000)    &   $\delta$ (J2000)  &  $R_{\rm e}^{a}$ [arcsec] & $M_{\rm r}$ [mag] & $M_{\rm \star}$ ($\times$ 10$^{10}$) [M$_\odot$]&   $\epsilon$  &  $\lambda_{Re}$\\
\hline
\hline
NGC0216 & Sd & 00 41 27.16  &   -21 02 40.82   & 20 & -18.99 & 0.19 & 0.711 & 0.52\\
NGC3057 & Sdm & 10 05 39    &   +80 17 12 & 32& -19.17 & 0.12 & 0.269 & 0.43 \\
NGC5682 & Scd & 14 34 44.97 &   +48 40 12.83& 26 & -19.39 & 0.25 & 0.764 & 0.68 \\
NGC7800 & Ir  & 23 59 36.75	&   +14 48 25.04 & 32 & -19.56 & 0.19 & 0.607 &  0.19 \\
UGC05990 & Sc & 10 52 37    &	+34 28 58 & 12 & -18.32 & 0.16 & 0.742 & 0.17\\
UGC08231 & Sd & 13 08 37.55 &	+54 04 27.73 & 19 & -19.28 & 0.14 & 0.664 & 0.40\\
UGC08733 & Sdm & 13 48 38.99&   +43 24 44.82 & 30 & -19.75 & 0.26 & 0.437 & 0.46\\
UGC10650 & Scd &17 00 14.58 &	+23 06 22.83 & 23 & -19.32 & 0.20 & 0.782 & 0.33\\
UGC10796 & Scd & 17 16 47.72&	+61 55 12.42 & 20 & -19.56 & 0.28 & 0.416 & 0.24\\
UGC12054 & Sc & 22 29 32.44	&   +07 43 33.68 & 15 & -18.41 & 0.10 & 0.739 & 0.41\\

\hline
\end{tabular}\\
Columns are: Name of target, morphological type, RA and DEC, effective radius, r-band absolute magnitude, stellar mass, ellipticity at 1$R_{\rm e}$ and specific angular momentum at 1$R_{\rm e}$. All the values were obtained from \cite{2019FalconBarroso}.\\

\end{table*}

\section{Instrumental resolution}
\label{Instruments_resoloution}
For measuring kinematics of a galaxy by fitting a stellar library, the data and the template library should have the same velocity scale. The latter is possible through convolution of the template spectra to the instrument's spectral resolution. To do so, the chosen library needs to have higher resolution than the observed data. However, the MUSE spectral resolution is not constant with wavelength, being lower in the blue part of the spectrum (4750 to 5500\,\AA\,) and higher in the red part (8400 to 9600 \AA) \citep{Krajnovic2015, 2015ApJ...804...70G, Mentz2016, Vaughan2018, 2019Emsellem}. The instrument resolution in the red part of the spectra is even higher than the spectral resolution of the E-MILES library. This makes the measurement of the velocity dispersion in low-surface-brightness galaxies, such as dEs, challenging as the velocity dispersion of the targeted dEs can be lower than the instrument resolution. This issue has already been mentioned in different studies of low-mass systems \citep[e.g., see][]{2004Emsellem, 2006Ganda, 2019Johnston,2019Emsellem}.

\begin{figure}
\centering
\includegraphics[scale=0.5]{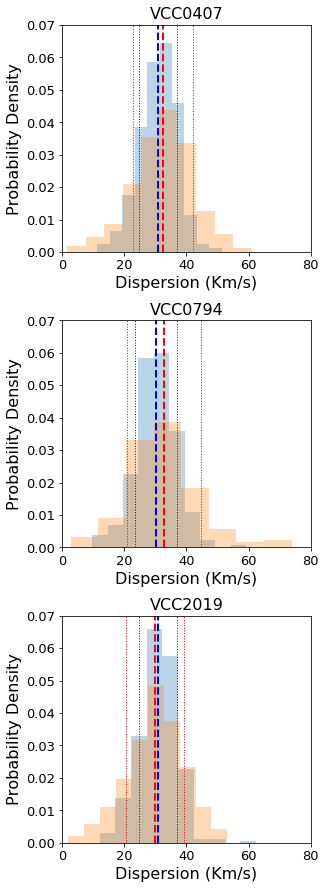}
\caption{The distribution of velocity dispersion measured with 1) the full spectrum fitting using the E-MILES library (in blue) and 2) the Ca triplet lines using the CaT library (in red) for three dEs in our sample. The mean value of each distribution is shown with a thick dashed line, while the thin dotted lines represent the standard deviation of the mean of the distribution. }
\label{CaTvsFull}
\end{figure}

Following \citet{2019Emsellem} and in order to test the robustness of our velocity dispersion measurements and their dependence on the library resolution, we repeated the fitting procedure by using the calcium II triplet (CaT) library \citep{cenarro2001}, encompassing the wavelength range of the Ca triplet lines, which has a higher resolution than E-MILES. The CaT library contains 706 stellar spectra covering the [Fe/H] range from $-$3.45 to 0.6 dex. This library is limited to the spectral range of 8350-9020 \AA\, with a FWHM $\sim$ 1.5 \AA\,\citep{cenarro2001}. Due to the small spectral coverage of the CaT library, we performed the pPXF fitting between 8400 and 9000 \AA. Adopting an instrument FWHM of 2.51 \AA\, we present the results of this test in Fig. \ref{CaTvsFull} for three dEs. In all panels, the distribution of the velocity dispersion as measured from each bin's average spectrum using E-MILES and the full spectrum fitting option is traced in blue, while the distribution of velocity dispersion derived with the CaT library is plotted in red. The mean value of each distribution is shown with a thick dashed line while the dotted lines of the same colour trace the $\pm$ 1 standard deviation of the distribution. 

As shown in Fig. \ref{CaTvsFull}, and consistent with the results of \cite{2019Emsellem} albeit for different sets of stellar libraries, fitting using both the E-MILES and CaT libraries retrieves a similar range of values for the velocity dispersion in three dEs of our sample. This means that the velocity dispersion of the stellar component obtained in this study from the full spectrum fitting is reliable and is not particularly affected by the instrument or library resolution. This is true for six dEs in our sample. The remaining three dEs have lower SNR due to shorter exposure time (VCC1833) or extended size (VCC 1836 and 0794) that cannot be fully mapped using MUSE. For these reasons, the sky background was not properly modeled and subtracted, thus hampering the quality of the fits in the CaT region \citep[for the effect of sky residuals also check][]{2019Johnston}. As an additional test, we used the varying resolution that mimics the MUSE instrument resolution, introduced by \cite{2015ApJ...804...70G} and \cite{2017Bacon}. Results of this test showed no particular difference in comparison to those obtained by using a fixed resolution of FWHM = 2.51 \AA\,
\citep{2010Bacon}

\section{Error Maps}
\label{Error_maps}
In Fig. \ref{errormaps_1}, \ref{errormaps_2}, and \ref{errormaps_3}, we provide error maps of velocity and velocity dispersion. Errors are measured through Monte-Carlo simulations, as described in Section \ref{Results}.

\begin{figure*}
    \centering
    \includegraphics[scale=0.7]{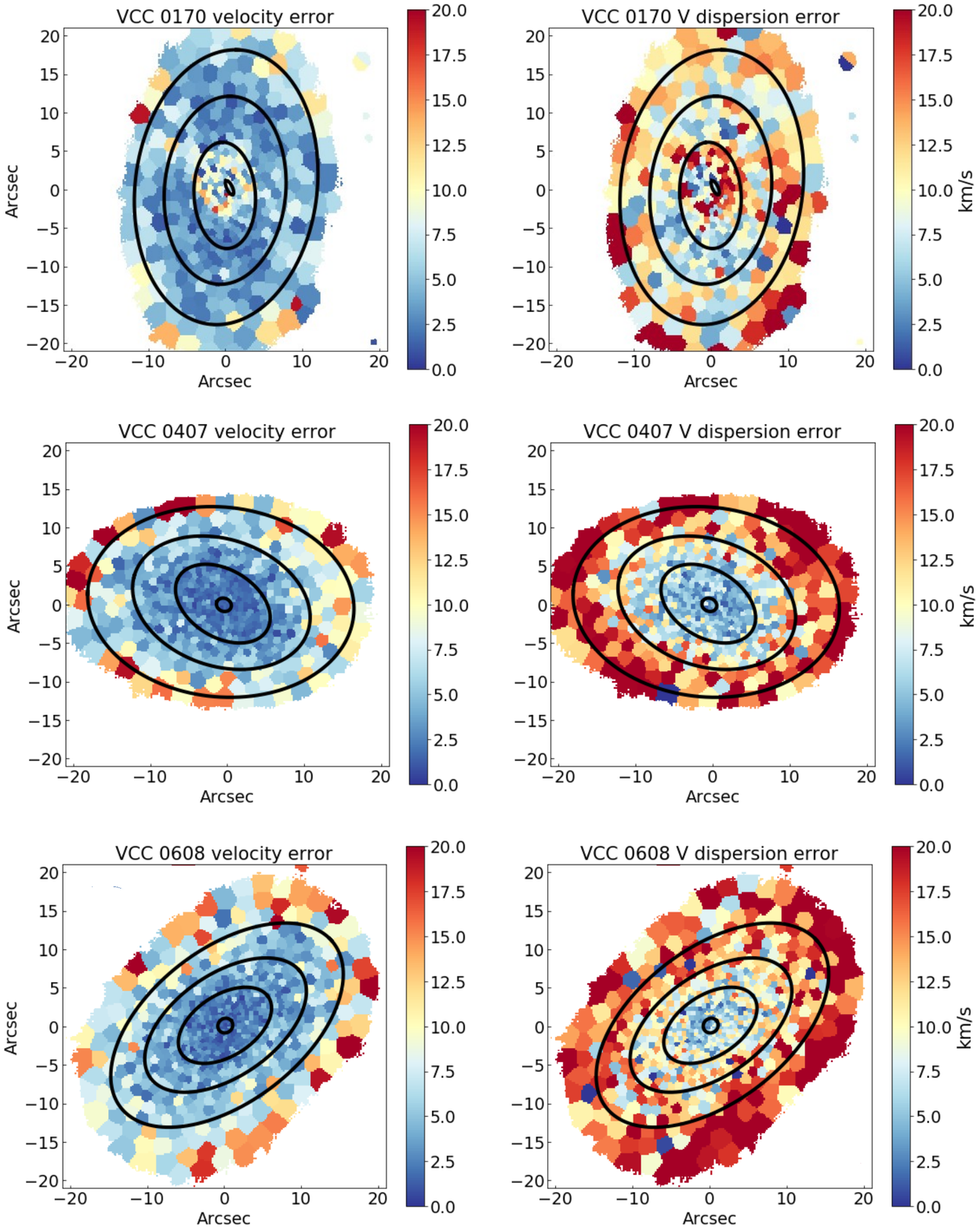}
    \caption{Error maps of velocity and velocity dispersion for our sample of dEs. For a better comparison, the isophotes from Fig. \ref{set1} and \ref{set2} are repeated in all the panels here. }
    \label{errormaps_1}
\end{figure*}

\begin{figure*}
    \centering
    \includegraphics[scale=0.7]{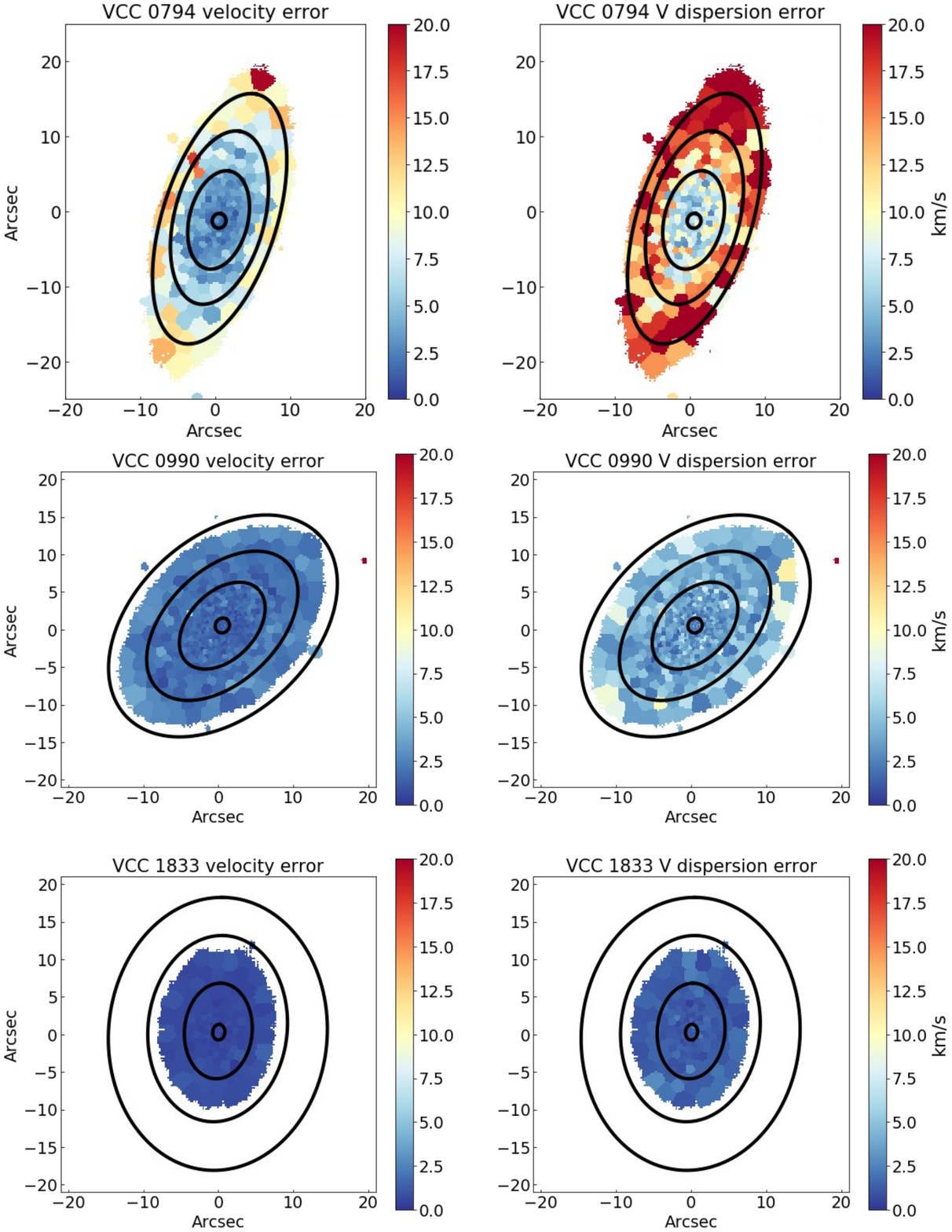}
    \caption{Continued}
    \label{errormaps_2}
\end{figure*}

\begin{figure*}
    \centering
    \includegraphics[scale=0.7]{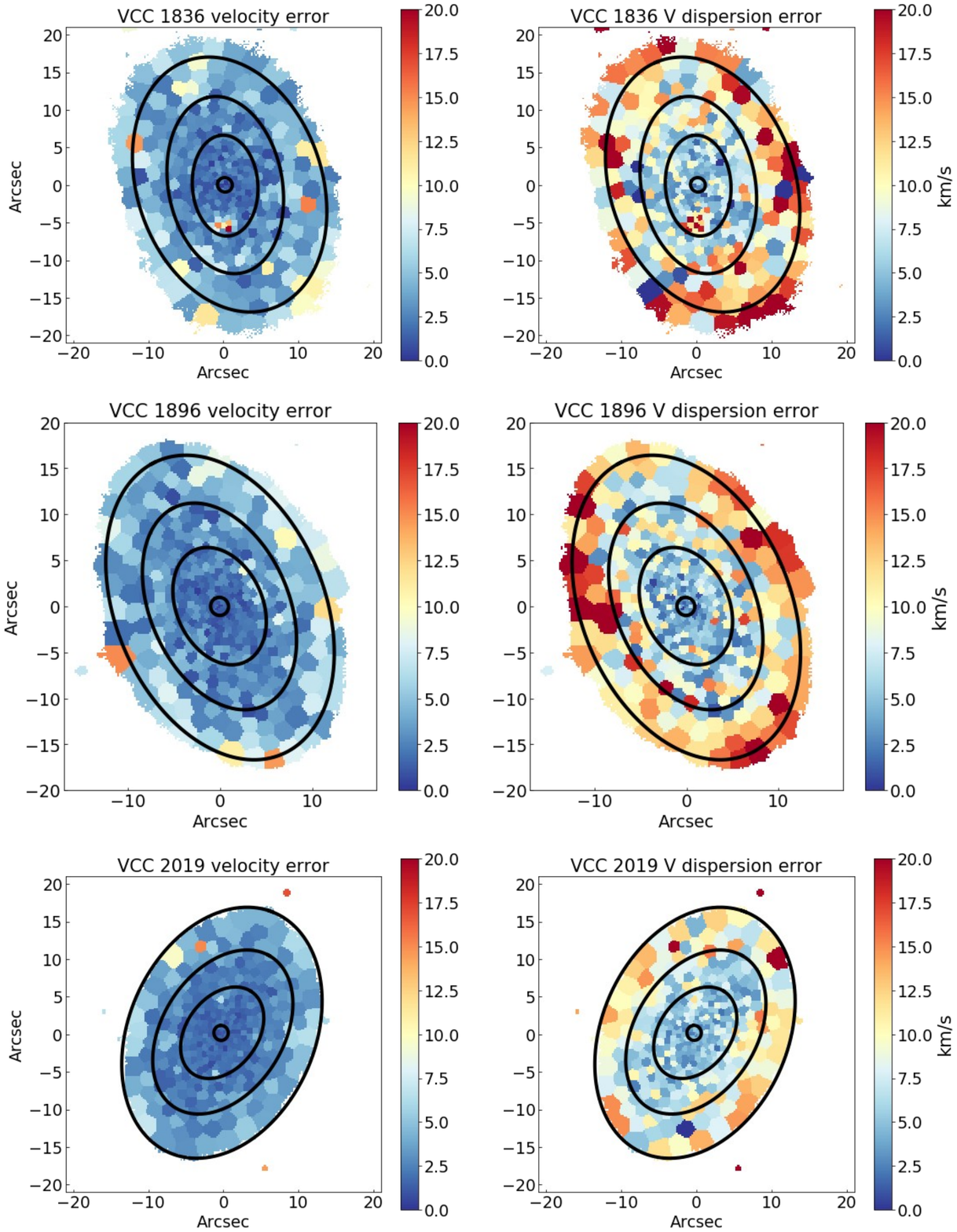}
    \caption{Continued}
    \label{errormaps_3}
\end{figure*}

\section{Emission lines in the center of VCC0170}
\label{VCC0170_emissionlines}

\begin{figure*}
\centering
\includegraphics[scale=0.7]{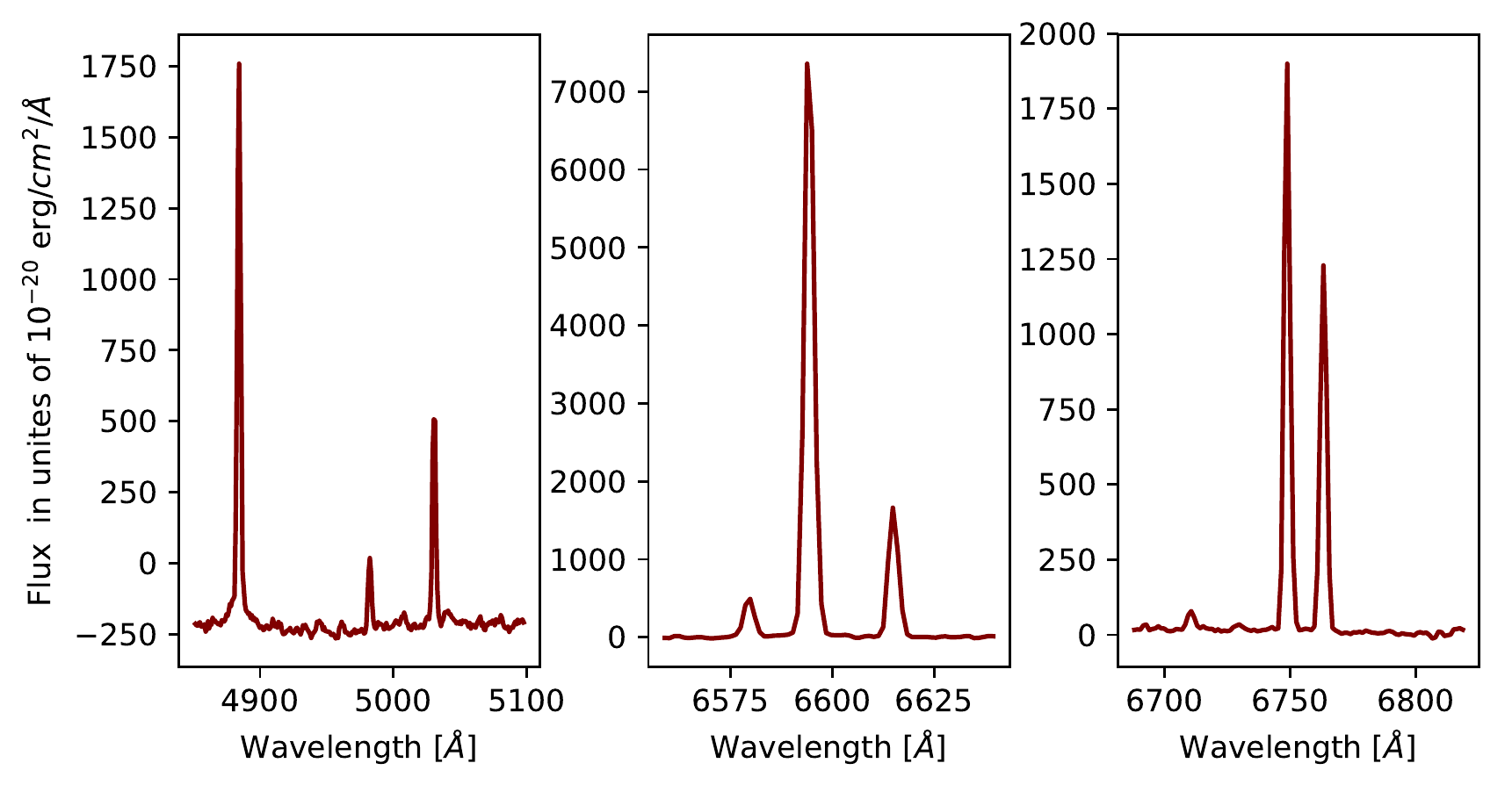}
    \caption{The observed nebular spectrum of the central region in VCC0170, corrected for the underlying galaxy emission. From left to right, the first panel shows: H$\beta$, [OIII]$\lambda$ 4958, and [OIII]$\lambda$ 5007; the second panel: [NII]$\lambda$ 6548, H$\alpha$, [NII]$\lambda$ 6583, and the third shows: [SII]$\lambda$ 6717 and [SII]$\lambda$ 6731.}
\label{VCC0170Emission}
\end{figure*}

As mentioned in Section \ref{VCC0170}, nebular emission lines were detected in the center of VCC0170. We summed all the MUSE spaxels within the central region of this galaxy (as shown in Fig. \ref{VCC0170Kin}), modeled the emission of the underlying stellar population with pPXF, and finally subtracted this model to obtain the nebular optical emission lines of H$\beta$, [OIII] $\lambda$ 4959, [OIII] $\lambda$ 5007, [NII]$\lambda$ 6548, H$\alpha$, [NII]$\lambda$ 6583, [SII] $\lambda$ 6717, and [SII] $\lambda$ 6731. These emission lines are shown in Fig. \ref{VCC0170Emission}. We corrected the H$\beta$ and H$\alpha$ fluxes for Galactic foreground extinction using $A_{v}= 0.089$ mag from \cite{2011Schlafly}. We also used the following Balmer decrement to correct the nebular emission lines for intrinsic reddening:

\begin{equation}
\\E(B-V)= log (f(H_{\alpha})/\rm f(H_{\rm \beta}) \times 2.85))\times (0.4(\rm \kappa_{a} -\rm \kappa_{b}))^{-1}
\end{equation}

where f(H$_{\rm \alpha}$) and f(H$_{\rm \beta}$) are the observed H$_{\alpha}$ and H$_{\beta}$ fluxes while $\kappa_{a}$ and $\rm \kappa_{b}$ are defined based on equations (3a) and (3b) of \cite{1989Cardelli}. We derived E(B-V) = 0.018 mag. In both cases, corrections were carried out using extinction law of \cite{1989Cardelli}. We obtained a  H$\alpha$ luminosity of 1.75 $\times$ 10$^{38}$ W. The observed fluxes of the identified emission lines are listed in Table \ref{VCC0170table}.

\begin{table}
	\caption{List of optical emission lines and their observed fluxes in VCC0170}
	\begin{threeparttable}
	\centering
	\begin{tabular}{c c}
	\hline
	Emission Line & Observed Flux \\
             	  & $\times{10^{-15}} (\rm erg s^{-1} cm^{-2})$\\
	\hline
	\hline
	H$_{\alpha}$             & 4.76\\ 
	H$_{\beta}$              & 1.64\\
	$[\rm OIII]$  $\rm \lambda$ 4959     & 0.20\\
	$[\rm OIII]$  $\rm \lambda$ 5007    & 0.66\\ 
	$[$NII$]$ $\rm \lambda$ 6548  & 0.35\\
	$[$NII$]$ $\rm \lambda$ 6583  & 1.04\\
	$[$SII$]$ $\rm \lambda$ 6717  & 1.12\\
	$[$SII$]$ $\rm \lambda$ 6731  & 0.80\\
	\hline
	\end{tabular}
	\label{VCC0170table} 
	\begin{tablenotes}
		\small
		\item Columns are: Name of nebular emission line, observed flux.
    \end{tablenotes}
	\end{threeparttable}
\end{table}

\section{Newly discovered Galaxies in the MUSE cubes}
\label{BB1_BB2}

\setlength{\tabcolsep}{1.5pt} 
\begin{table}
	\caption{New galaxies in the MUSE cubes: BB1 and BB2}
	\begin{threeparttable}
		\centering
		\begin{tabular}{c c c c c}
		\hline
		Name           & RA (J2000)   &  Dec (J2000)   &  redshift&   total $M_{V,AB}$ \\
		\hline
		\hline
		BB1            & 12 40 19.50 & +14\ 43\ 0.0 & 0.5528 $\pm$ 0.0002 &         21.32           \\
		BB2            & 12 14 20.50 & +14\ 26\ 2.0 & 0.349 $\pm$ 0.013   &         21.50 \\
		\hline
		\end{tabular}
		\label{newgalaxies}
		\begin{tablenotes}
		\small
		\item Columns are: Name of the target, RA, DEC, redshift and total AB magnitude in V band.
		\end{tablenotes}
	\end{threeparttable}
\end{table}

\begin{figure*}
\centering
\includegraphics[scale=0.43]{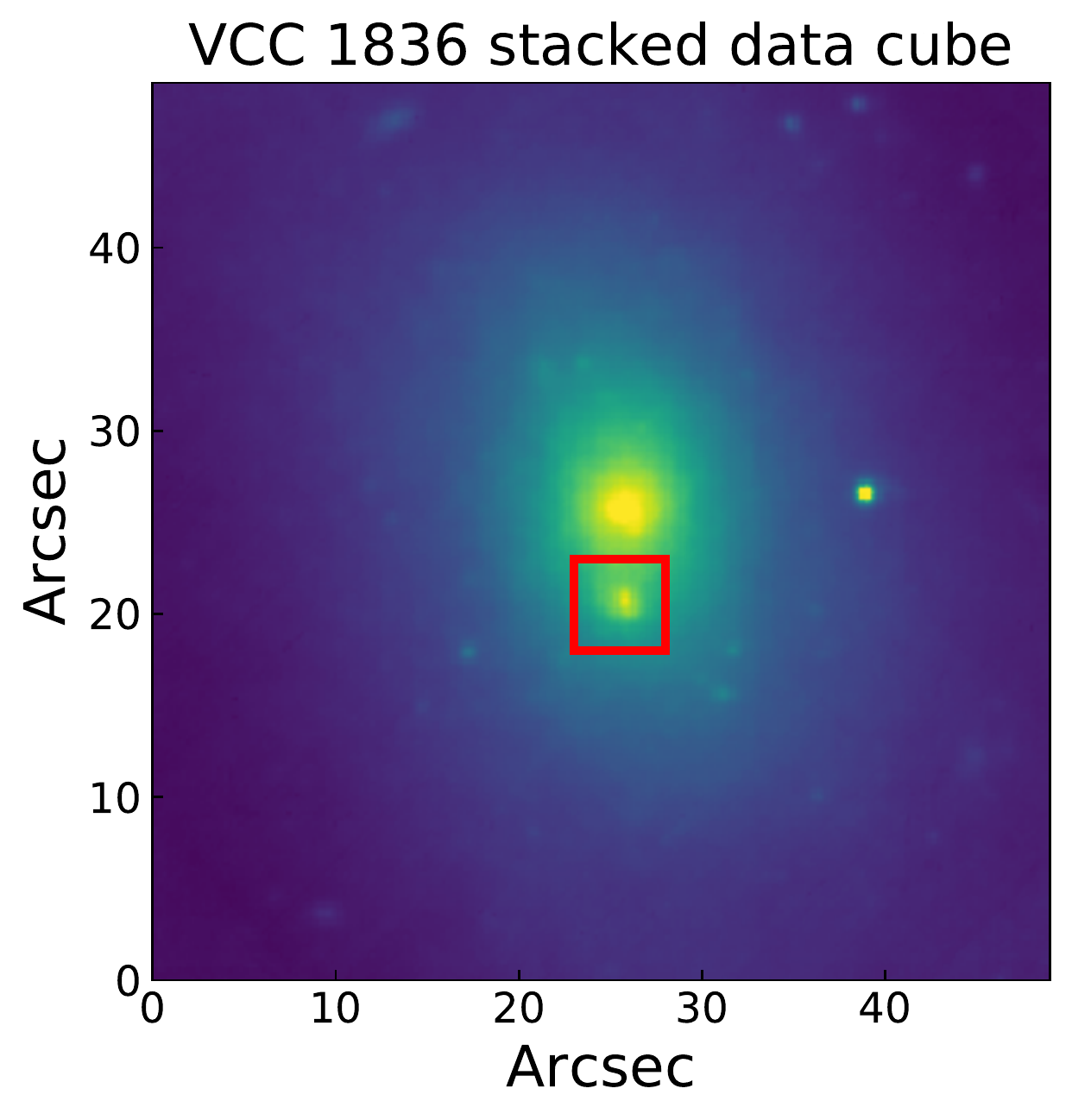}
\includegraphics[scale=0.4]{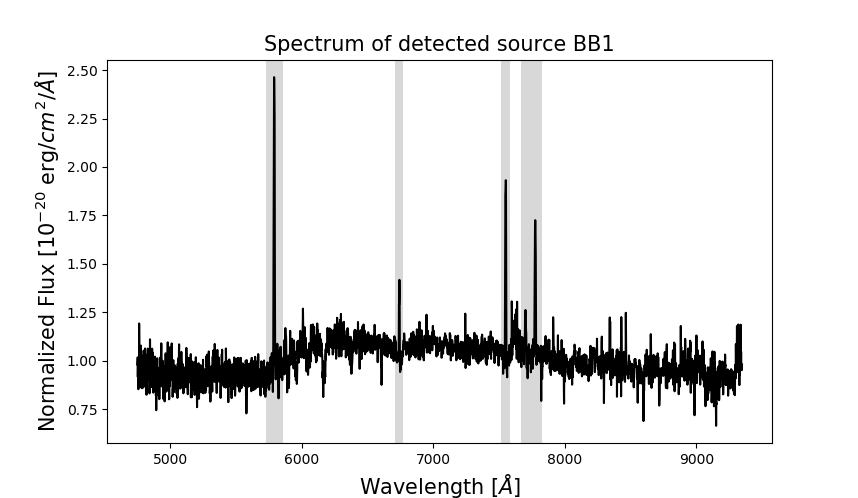}
\caption{A source detected in the projected vicinity of the central region of VCC1836. Left panel: The location of the detected source (BB1) is marked with a red box with a size of $8 \times 8$ arcsec$^2$ on the MUSE stacked image. Right panel:The observed spectrum of the detected source, corrected for the underlying emission of VCC1836. The following emission lines are marked from left to right: O[II]$\lambda$ 3727, H$\gamma$, H$\beta$, [OIII]$\lambda$ 4958, [OIII]$\lambda$ 5007. }
\label{BB1}
\end{figure*}

As mentioned in Section \ref{Results}, nebular emission lines were detected in a particular region inside VCC 1836 in the vicinity of the dE's center. Apart from the bright nucleus of VCC1836, the SDSS images of this dE also show a faint light concentration in the central region of this galaxy. Thanks to the high spatial resolution of MUSE, this detected source is better resolved in the MUSE stacked images of our dataset, as shown in Fig. \ref{BB1} (red box). We extracted the spectrum of this particular region (hereafter BB1) within an aperture with a radius of 5 pixels, corresponding to an area of 3 arcsec$^2$. We subtracted the possible light contamination of VCC 1836 by subtracting a spectrum extracted within an annulus of the same size around the detected source. The final spectrum is plotted in the right panel of Fig. \ref{BB1}. We measured the redshift of this background galaxy, using the Doppler shift of each nebular emission line. Based on the average redshift, this galaxy is located at z = 0.5528 $\pm$ 0.0002 (also reported in Table \ref{newgalaxies}).
We modeled the continuum of this spectrum using pPXF in order to subtract the emission of the underlying stellar population and to measure the flux of the nebular emission lines. The values are reported in the second column of Table \ref{emissiontable}.  

\begin{figure*}
\centering
\includegraphics[scale=0.45]{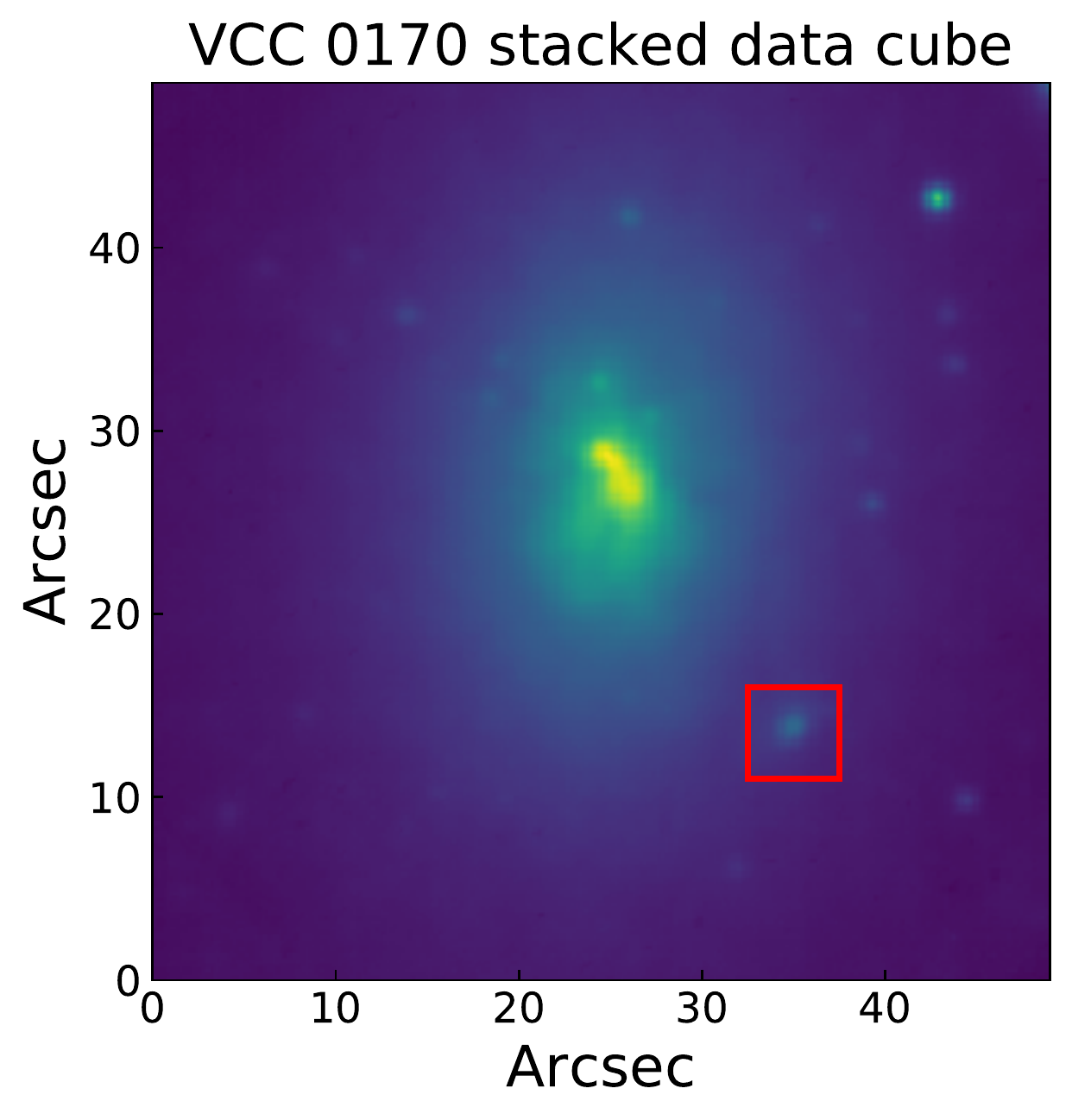}
\includegraphics[scale=0.4]{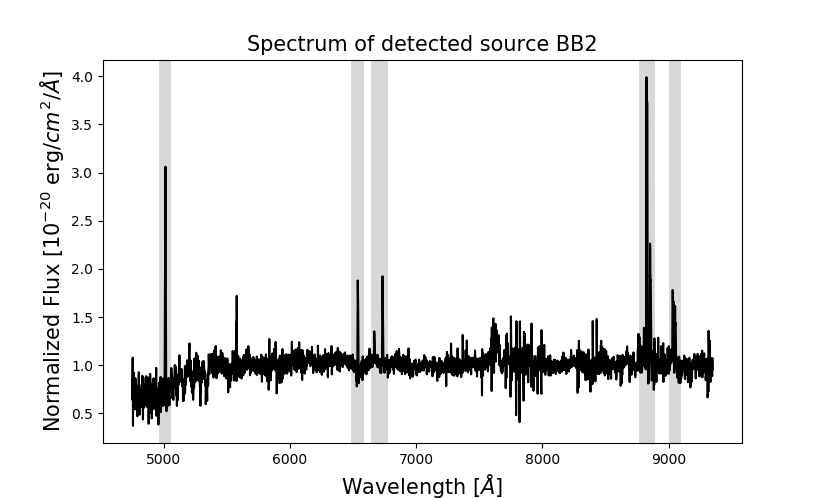}
\caption{Detected background source in the FOV of VCC0170. Left panel: The location of the detected source (BB2) is marked with a red box with a size of $8 \times 8$ arcsec$^2$ on the MUSE stacked image. Right panel: The observed spectrum of the detected source, corrected for the underlying emission of VCC0170. Marked emission lines from left to right: [OII]$\lambda$ 3727,  H$\beta$, [OIII]$\lambda$ 4958, [OIII]$\lambda$ 5007, H$\alpha$, [NII]$\lambda$ 6548, [NII]$\lambda$ 6583, [SII]$\lambda$ 6717, and [SII]$\lambda$ 6731.}
\label{BB2}
\end{figure*}

Another source with strong emission lines was detected in the MUSE cube of VCC 0170 (hereafter BB2). In Fig. \ref{BB2} we marked the location of this extended source with a red box. As in the previous case, we extracted the spectrum of this source within a radius of 7 pixels, corresponding to an area of 6 arcsec$^2$. The sky and the VCC 0170 emission were removed by subtracting a spectrum extracted within an aperture of the same size near the source. The resulting spectrum is plotted in the right panel of Fig. \ref{BB2}. We measured the redshift of this source using the nebular emission lines. This background galaxy is at a redshift $z$ = 0.349 $\pm$ 0.013. We modeled the continuum and absorption features using pPXF and subtracted the model from the source spectrum to measure the line fluxes. They are reported in the third column of Table \ref{emissiontable}.

To locate this galaxy in the BPT diagram \citep{1981Baldwin}, we corrected the emission lines of H$\alpha$, [OIII]$\lambda$ 5007, H$\beta$ and [NII] for Galactic foreground extinction, using $A_{v} = 0.089$ mag as reported by \cite{2011Schlafly}. The same emission lines were also corrected for intrinsic reddening using the Balmer decrement as described in Section \ref{VCC0170_emissionlines} (E(B-V)= 0.055 mag). For both corrections we used the extinction law of \cite{1989Cardelli}. We also corrected the observed flux for redshift dimming following \cite{2014Calvi}:

\begin{equation}
\\I = I_{0} (1+z)^{-4}
\end{equation}

where I and I$_{\rm 0}$ are the observed and intrinsic surface brightness. 
Following the corrections above, we measured log[$[\rm OIII]$ $\lambda$ 5007/{\rm H$\beta$}] = $-0.038$ and log[$[\rm NII]$ $\lambda$ 6584/{\rm H$\alpha$}] = $-0.351$. According to these values, the detected galaxy is located at the edge of the star forming galaxies in the BPT diagram, as defined by the line of \cite{2003Kauffmann}. We used the measured redshift to estimate the co-moving radial distance, which turns out to be $\approx$ 1360.4 Mpc. Using this value, we measured the star formation rate of this background galaxy to be 0.66 M$_{\odot}$ yr$^{-1}$, following the method of \cite{2007Calzetti}.

\begin{figure*}
\centering
\includegraphics[scale=0.4]{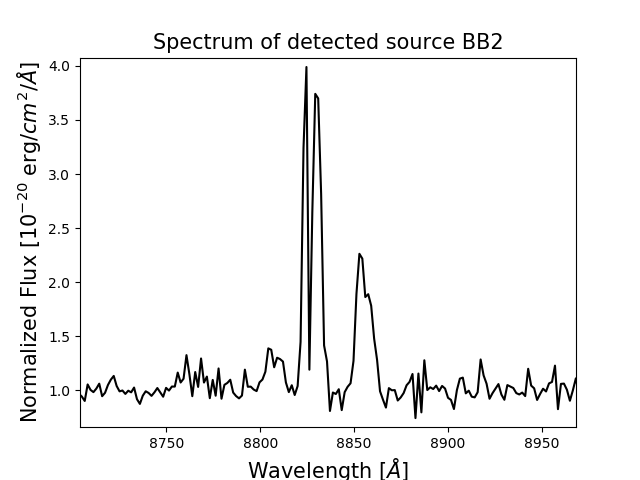}

\caption{Split H$\alpha$ and [NII] 6583 emission lines in the red part of the spectrum of BB2. This split corresponds to a rotation velocity of $90 \pm 10$ km\,s$^{-1}$. }
\label{BB2_spect}
\end{figure*}

Further analysis of the gas emission lines present in the spectrum of this galaxy shows that the H$\alpha$, [SII], and [NII] lines are split (Fig. \ref{BB2_spect}). This can likely be due to the rotation of this background galaxy. We used the split in the lines and measured the rotation of this galaxy to be of the order of $\approx$ 90 km\,s$^{-1}$, with an error of approximately 10 km\,s$^{-1}$. 

\begin{table*}
\caption{\label{emissiontable} List of optical emission lines and their observed fluxes in BB1 and BB2}
\centering
\begin{tabular}{c c c}
\hline
Emission       & Observed Flux in BB1    &  Observed Flux in BB2\\
Lines       & $\times{10^{-17}} (\rm erg s^{-1} cm^{-2})$ &  $\times{10^{-17}} (\rm erg s^{-1} cm^{-2})$\\
\hline
\hline
H$_{\alpha}$             & --    & 15.8\\ 
H$_{\gamma}$             & 3.62  & --  \\ 
H$_{\beta}$              & 7.43  & 51.5\\
$[\rm OII]$                  & 14.7  & 17.6    \\
$[\rm OIII]$ $\rm \lambda$ 4959    & 1.68  & 1.66  \\
$[\rm OIII]$ $\rm \lambda$ 5007    & 5.64  & 4.86 \\
$[\rm NII]$ $\rm \lambda$ 6548     & 0.35  & 2.33\\
$[\rm NII]$ $\rm \lambda$ 6583     & 1.04  & 7.08\\
\hline
\end{tabular}\\
\end{table*}

\end{appendix}

\label{lastpage}
\end{document}